\documentclass{emulateapj}
\slugcomment{Submitted to {\it ApJ} on 05 May 2008; Accepted on 04 Jun 2008}
\shortauthors{Looper et al.}
\shorttitle{Two Peculiar L Dwarfs}

\begin{document}

\title{Discovery of Two Nearby, Peculiar L Dwarfs from the 2MASS Proper
Motion Survey: Young or Metal-Rich?\altaffilmark{1}}

\author{Dagny L. Looper\altaffilmark{2,3}, 
J. Davy Kirkpatrick\altaffilmark{3,4}, 
Roc M. Cutri\altaffilmark{4}, 
Travis Barman\altaffilmark{5}, 
Adam J. Burgasser\altaffilmark{3,6}, 
Michael C. Cushing\altaffilmark{2,7}, 
Thomas Roellig\altaffilmark{8}, 
Mark R. McGovern\altaffilmark{9}, 
Ian S. McLean\altaffilmark{10}, 
Emily Rice\altaffilmark{10}, 
Brandon J. Swift\altaffilmark{7}, 
Steven D. Schurr\altaffilmark{4}}

\altaffiltext{1}{Based in part on data collected at 
Subaru Telescope, which is operated by the National Astronomical Observatory 
of Japan.  Some of the data presented herein were obtained at the W.M. 
Keck Observatory, which is operated as a scientific partnership among the 
California Institute of Technology, the University of California and the 
National Aeronautics and Space Administration. The Observatory was made 
possible by the generous financial support of the W.M. Keck Foundation.}

\altaffiltext{2}{Institute for Astronomy, University of
Hawai'i, 2680 Woodlawn Dr, Honolulu, HI 96822} 

\altaffiltext{3}{Visiting Astronomer at the Infrared Telescope Facility, which
is operated by the University of Hawaii under Cooperative Agreement
no. NCC 5-538 with the National Aeronautics and Space Administration,
Office of Space Science, Planetary Astronomy Program} 

\altaffiltext{4}{Infrared Processing and 
Analysis Center, MS 100-22, California
Institute of Technology, Pasadena, CA 91125} 

\altaffiltext{5}{Lowell Observatory, 1400 W. Mars
Hill Rd., Flagstaff, AZ 86001} 

\altaffiltext{6}{MIT Kavli Institute for Astrophysics \&
Space Research, 77 Massachusetts Ave, Building 37-664B, Cambridge, MA
02139}

\altaffiltext{7}{Steward Observatory, University of 
Arizona, 933 N. Cherry Ave., Tucson, AZ 85721} 

\altaffiltext{8}{NASA AMES, 7.  Research Center, Moffett Field, 
CA 94035-1000} 

\altaffiltext{9}{Antelope Valley College, 3041 West Avenue K, 
Lancaster, CA 93536} 

\altaffiltext{10}{Department of Physics and Astronomy, University 
of California, Los Angeles, CA 90095-1562}

\begin{abstract}

We present the discovery of two nearby L dwarfs from our 2MASS 
proper motion search, which uses
multi-epoch 2MASS observations covering $\sim$4700 square degrees
of sky.  2MASS J18212815+1414010 and 2MASS J21481628+4003593 
were overlooked by
earlier surveys due to their faint optical magnitudes and their
proximity to the Galactic Plane (10$^{\circ}$ $\le$ $|$b$|$ $\le$
15$^{\circ}$).  
Assuming that both dwarfs are single, we derive spectrophotometric 
distances of $\sim$10 pc, thus increasing the number of 
known L dwarfs within 10 pc to 10.  In the near-infrared, 
2MASS J21481628+4003593 shows a triangular-shaped $H$-band
spectrum, strong CO absorption, and a markedly red $J-K_s$ color
(2.38~$\pm$~0.06) for its L6 optical spectral type.
2MASS J18212815+1414010 also shows a
triangular-shaped $H$-band spectrum and a slightly red $J-K_s$ color
(1.78~$\pm$~0.05) for its L4.5 optical spectral type.  
Both objects show strong silicate absorption at 9--11 $\mu$m.  
Cumulatively, these features imply an unusually dusty photosphere for 
both of these objects.  We examine several scenarios to explain the 
underlying cause for their enhanced dust content  
and find that a metal-rich atmosphere or a low-surface gravity   
are consistent with these results.  2MASS J18212815+1414010 may be
young (and therefore have a low-surface gravity) 
based on its low tangential velocity of 10 km s$^{-1}$.  On the
other hand, 2MASS J21481628+4003593 has a high tangential velocity of 62
km s$^{-1}$ and is therefore likely old.  Hence, high metallicity and
low-surface gravity may lead to similar effects.

\end{abstract}


\section{Introduction}

Many of the objects known to reside in the Solar Neighborhood were
discovered via their high proper motion. Early proper motion
surveys by the pioneers of photographic astrometry
built the foundation of our knowledge about the local stellar census
(e.g., \citealt{luyten1979, luyten1980, giclas1978, ross1939, 
wolf1918, barnard1916, innes1915}).  Discoveries of our nearest stellar
and substellar neighbors continue to-date, with surveys such as 
the SIPS survey \citep{deacon2007, deacon2005}, 
the LEHPM survey \citep{pokorny2004}, the SCR survey \citep{finch2007},
and the SUPERBLINK survey \citep{lepine2008b, lepine2008a, lepine2002}.
Despite their successes, these surveys are not ideally suited to the  
discovery of nearby brown dwarfs because of their dependence on  
optical data.  As we have now learned from longer-wavelength surveys such
as the Two Micron All Sky Survey (2MASS; \citealt{skrutskie2006}), 
the Deep Near Infrared Survey of the southern sky 
(DENIS; \citealt{epchtein1997}), 
and the Sloan Digital Sky Survey (SDSS; \citealt{york2000}), the
Solar Neighborhood is home to many L and T dwarfs -- the coldest brown
dwarfs currently known \citep{kirkpatrick2005}. 
Most of these cool
objects have been selected photometrically using color selections 
appropriate for L and T dwarfs with near solar metallicity. This method 
may be missing low-temperature objects with unusual physical properties
-- low or high metallicity dwarfs or objects with unusual atmospheric
traits --  but a 
measurement of their high motions will help to identify them as nearby.

Another benefit of proper motion surveys over photometric ones is their 
ability to find nearby objects against the confusion of the Galactic
Plane.  Due to reddening caused by intervening material, extinguished 
background objects in the Plane can have colors mimicking those of 
nearby dwarfs in 2MASS, DENIS, and SDSS colors. Proper motions help 
ferret out these nearby interlopers without relying on color information.

Using data contained in the 2MASS Survey Working Databases, we have 
searched for proper motion objects by examining the subset of
multi-epoch data having time differences in excess of two months. The 
total coverage satisfying this criterion is approximately 4700 square 
degrees, or $\sim$11\% of the sky. In this paper we present  
two discoveries from this exclusively near-infrared search. Both objects are 
optically faint, nearby L dwarfs located within 15$^\circ$ of the 
Galactic Plane and hence missed by previous optical proper motion and 
near-infrared photometric surveys.  We briefly describe our 2MASS proper motion
search and measurements from the 2MASS catalog in $\S$2, along with the
observations; we give the analysis in $\S$3; discussion in $\S$4; and
conclusions in $\S$5.

\section{Observations}

\subsection{Discovery}

We have conducted a near-infrared (NIR) proper motion search using the 
2MASS Database for high 
proper motion and late-type, optically faint objects 
\citep{burgasser2007a,looper2007a}.  2MASS was designed 
to be a single epoch survey covering the entire sky, but for various quality 
control or scientific reasons, multiple regions of the sky were observed 
more than once, totalling $\sim$4700 sq degrees of sky observed over at least 
two epochs.  We briefly describe our search criteria here; a complete
description of the survey will be presented in a forthcoming publication 
(Kirkpatrick et al$.$, in prep).

Candidates were required to have $J$~$\le$~15.8, have no optical counterparts 
($B$- and $R$-bands) within 5$\arcsec$ in the USNO-B1 Catalog
\citep{monet2003}, have positional differences of 
0$\farcs4~\le~\Delta$x$~\le~1\farcs$5 between the first and last epochs, 
have first and last epoch time differences of $\Delta$t~$\ge$~0.2 yr, 
proper motions of $\mu~\ge~0\farcs$2 yr$^{-1}$, and be located at least 
10 degrees away from the Galactic Plane.  After this initial query, 
candidates were examined visually using charts constructed 
from 2MASS $J$, $H$, and $K_s$ bands; Digitized Sky Survey (DSS) I
$R$, $B$, or $V$-band; and DSS II $R$-band.  
DSS I and II images were obtained from the CADC 
service\footnote{See http://cadcwww.dao.nrc.ca/cadcbin/getdss.}. 
Those without visible counterparts in the DSS I or II images or, 
for particularly bright objects with large $R-J$ colors ($R-J~\gtrsim~6$), 
were selected for follow-up spectroscopy.   
Here we present two noteworthy objects selected for early follow-up 
based on their bright 2MASS magnitudes.

\subsection{Two Bright L Dwarf Candidates}

\begin{figure}
\epsscale{0.9}
\plotone{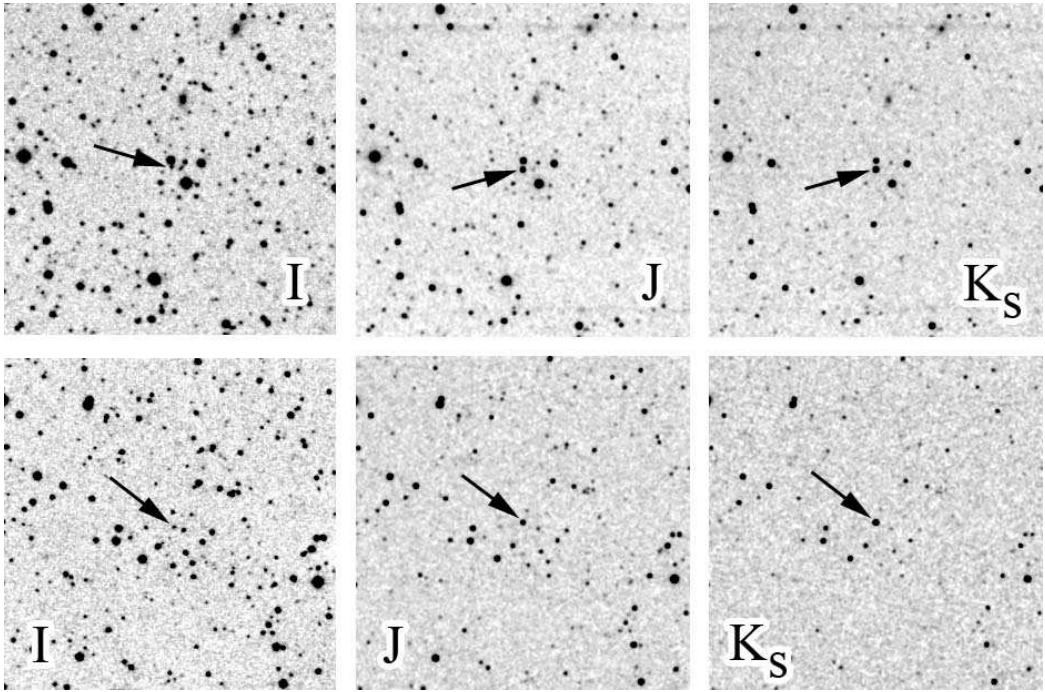}
\caption{Finder charts for 2MASS J18212815+1414010 (upper row) and
2MASS J21481628+4003593 (lower row). Each field is centered
on the target and is 5 arcminutes wide with North up and East to the left.
From left to right the three images for 2MASS J1821+1414 are $I$-band from
the DSS II (epoch 29 Jun 1993), $J$-band from 2MASS (epoch 27 Mar 1999),
and $K_s$-band from 2MASS (epoch 27 Mar 1999). For 2MASS J2148+4003 the images are
$I$-band from DSS II (epoch 22 Sep 1993), $J$-band from 2MASS (epoch 30
May 2000) and $K_s$-band from 2MASS (epoch 30 May 2000). Arrows mark the location of the
objects on the DSS II and 2MASS images.
\label{finders}}
\end{figure}

The first candidate has a 2MASS All Sky Point Source Catalog designation
of 2MASS J18212815+1414010, hereafter abbreviated 2MASS J1821+1414.
Astrometric and photometric measurements from the two separate epochs 
listed in the 2MASS Survey Point Source Working Database are given in Table 1.  
Despite this object having bright $J$-band (13.43~$\pm$~0.03) and $K_s$-band
(11.65~$\pm$~0.03) magnitudes, it is invisible in the
DSS I $B$-band and DSS II $R$-band plates.  It is, however, visible in
the DSS II $I$-band plate (see Figure~\ref{finders}).  Its large implied 
optical-to-infrared color along with a $J-K_s$ color of 1.78~$\pm$~0.04 
suggest that it is a mid-to late-type L dwarf. Using the two available
2MASS epochs and a positional error estimate of 0${\farcs}$06, 
we calculate a proper motion of 
$\mu$~=~0${\farcs}$21~$\pm$~0${\farcs}$07 at PA=153.4$^\circ$~$\pm$~0.7$^\circ$.

\begin{figure}
\epsscale{0.8}
\plotone{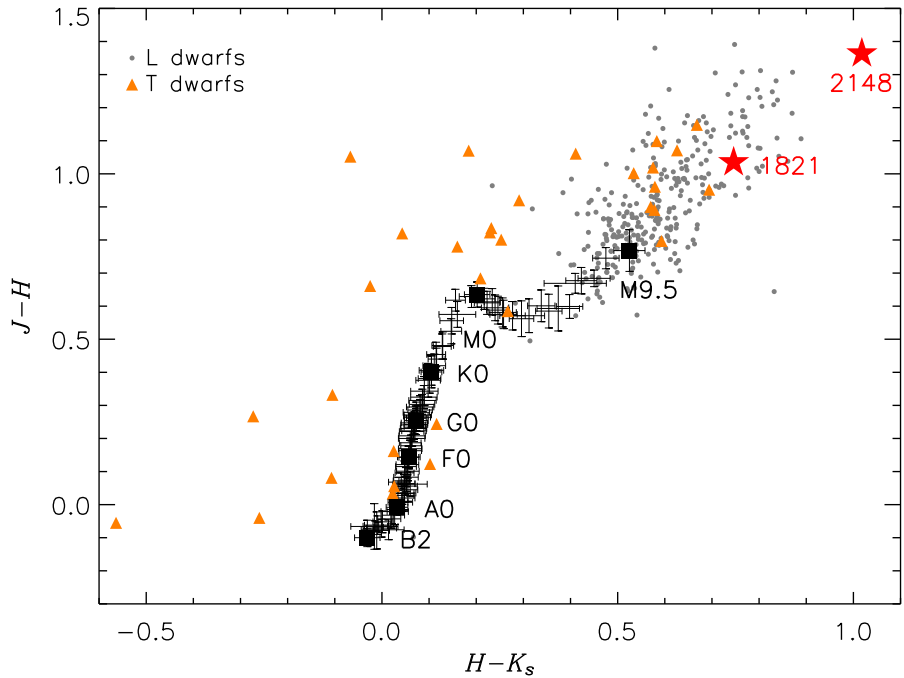}
\caption{Near-infrared color-color plot ($H-K_s$ vs $J-H$)
showing (1) average colors of B2--M9.5 dwarfs in the 2MASS Point Source
Catalog with each half subtype represented by a point with its
associated error bar and average colors for labeled spectral types
indicated by solid black squares; (2) L dwarfs with 2MASS 
photometry (see http://dwarfarchives.org) 
and $\sigma_{J-H}$ \&
$\sigma_{H-K_s}$ $\le$ 0.1 are shown as filled gray circles; 
(3) T dwarfs with 2MASS photometry
and $\sigma_{J-H}$ \& $\sigma_{H-K_s}$ $\le$ 0.2 are shown as solid
orange triangles; (4) the positions of 2MASS J1821+1414 and 
2MASS J2148+4003 are marked by solid red five-pointed stars.
\label{NIRcolors}}
\end{figure}

The second candidate has a designation of 2MASS J21481628+4003593, 
hereafter abbreviated as 2MASS J2148+4003. Measurements of this object 
from the 2MASS Survey Point Source Working Database are also given in 
Table 1. 2MASS J2148+4003 has a very red $J-K_s$ color (2.38~$\pm$~0.06), 
bright $J$-band (14.15~$\pm$~0.04) and $K_s$-band 
(11.77~$\pm$~0.04) magnitudes, and is invisible in the DSS I 
$R$-band and DSS II $R$-band plates.  
It is visible on the DSS II $I$-band plate (see Figure~\ref{finders}).  
Using the two available 2MASS epochs
and a positional error estimate of 0${\farcs}$06, 
we calculate a high proper motion of 
$\mu$~=~1{\farcs}33~$\pm$~0{\farcs}24 at PA=75.6$^\circ$~$\pm$~0.3$^\circ$. 
We show the position of both 2MASS J1821+1414 and 2MASS J2148+4003 on a NIR
color-color diagram along with the average colors of main sequence dwarfs
and L and T dwarfs with 2MASS photometry in Figure~\ref{NIRcolors}.  
Both objects lie along the redward extension of late-type M to L dwarfs.

In the following sections we describe observations of 2MASS 
1821+1414 and 2MASS J2148+4003 carried out
using Subaru-Focas, Keck-NIRSPEC, IRTF-SpeX, and Spitzer-IRS 
along with data reductions.  The
summary for all observations can be found in Table 2.  

\subsection{Subaru-FOCAS Spectroscopy}

The optical spectroscopic observations were carried out on 
20 and 21 Aug 2007 (UT) at the Subaru Telescope on Mauna Kea, 
Hawai'i, using the Faint Object Camera And Spectrograph (FOCAS; 
\citealt{2002PASJ...54..819K}). FOCAS was used with the 
300R grating blazed at 7500 \AA~and the S058 filter to block 
second-order light from wavelengths shorter than 5800 \AA. The 
grating tilt was set so that the wavelength region from 5850 to 
10250 \AA~was covered. Use of a long slit of width $0{\farcs}8$ 
resulted in a resolution of 8.5 \AA. Both nights were clear but 
the seeing was variable over the course of both nights. On 20 
Aug 2007 the seeing varied from $0{\farcs}5$ to $2{\farcs}5$. 
On 21 Aug 2007 the seeing was somewhat more stable, ranging only 
from $0{\farcs}6$ to $1{\farcs}5$. Subaru employs an atmospheric 
dispersion corrector, so keeping the slit aligned with the 
parallactic angle was not necessary.

The data were reduced and calibrated using standard IRAF\footnote{IRAF 
is distributed by the National Optical
Astronomy Observatories, which are operated by the Association of
Universities for Research in Astronomy, Inc., under cooperative
agreement with the National Science Foundation.} routines. 
The overscan region of the array was used 
for bias subtraction, and a median of five dome flats taken on the 
first night was used to normalize the response of the detector. 
Because of strong, broad spectral signatures in the flat field 
lamps themselves, special care was needed during the flat field 
correction step. The FOCAS chip is designed to be used with multi-object 
spectroscopy, but for our single-slit observations most of the detector 
area is not used or needed for sky subtraction. This allowed us to 
perform a block average (in the spatial dimension) on a much smaller 
21$\times$3260 pixel region of the median dome flat corresponding to 
the area where the primary target spectrum and sky subtraction region 
would fall in subsequent exposures. This block-averaged,
one-dimensional, vertical slice encompasses the gross undulations of 
the dome flat in the spectral dimension, and we replicated this slice 
across the spatial dimension to create a two-dimensional image. This 
image was then divided back into the median flat to produce a map 
that contains only small-scale flat-field variations. This map was 
then normalized using IMSURFIT and divided into all other data frames 
to remove the flat field signature across the detector.

The individual stellar spectra were then extracted using the APALL 
package. Wavelength calibration was achieved using arc lamps taken 
after each program object. Spectra were flux calibrated using an 
observation of LTT 9491 (\citealt{1994PASP..106..566H}) taken on the first 
night. We checked this fluxing using another standard, Wolf 1346 
(\citealt{1990ApJ...358..344M}), taken on the second night and the 
agreement was excellent. Dwarfs of type G0 were also acquired after 
each program target, and these were used to correct for telluric 
absorption in the program objects. Specifically, the G0 
dwarf SAO 10356 was used for 2MASS J1821+1414 and the G0 dwarf SAO 
207827 was used for 2MASS J2148+4003. The telluric bands removed by 
this method are those of O$_2$ at 6867--7000 (the Fraunhofer B band) 
and 7594--7685 \AA~(the Fraunhofer A band) as well as H$_2$O at 
7186--7273, 8162--8282, and $\sim$8950--9650 \AA.

\subsection{Keck-NIRSPEC Spectroscopy}

2MASS J1821+1414 and 2MASS J2148+4003 were observed with the Near-Infrared
Spectrometer (NIRSPEC, \citealt{mclean1998,mclean2000}) on the 10-m 
W.~M.~Keck Observatory.  Using a 1024$\times$1024 InSb array and the 
spectrograph in low-resolution mode, we selected the N3 filter to
cover the part of the $J$-band window from 1.14 to 1.36 $\mu$m where the 
most diagnostic NIR features lie. Use of the $0{\farcs}38$
slit results in a resolving power of 
R~$\equiv~{\lambda}/{\Delta}{\lambda}~{\approx}~2500$.  For the observations of
2MASS J1821+1414 on 2005 July 18 UT seeing was 
1{\farcs}5, but sky transparency was otherwise fine.
Sky conditions during the 2005 Dec 09 UT observations of 2MASS
2148+4003 were excellent.  The data were obtained in two sets of 
dithered pairs, with a 300-second exposure obtained at each position and 
an airmass of 1.00 and 1.20 for 2MASS J1821+1414 and 2MASS J2148+4003, 
respectively. To measure telluric absorption, the A0 dwarf HD 165029 was 
observed for 2MASS J1821+1414 in one set (two nods) of 30 coadds of 
2-sec each and one set of 30 coadds of 1-sec each, at an airmass of 
1.00 and 1.01, respectively.  For 2MASS J2148+4003, the A0 dwarf SAO 
71693 was observed in two sets of dithered pairs with a 30-sec exposure 
obtained at each position and an airmass of 1.26.  
Such early-type A stars are essentially featureless in this region;
only the star's hydrogen Paschen line at 1.282 $\mu$m needing to be 
interpolated over during the telluric correction step. Calibrations 
consisted of neon and argon arc lamp spectra, a dark frame, and a
spectrum of a flat-field lamp.  We employed standard reductions using 
the REDSPEC package, as described in detail in \cite{mclean2003}.

2MASS J2148+4003 was again obtained with NIRSPEC on 2006 Aug 04 UT in
two more low-resolution settings covering the wavelength regions
0.94$-$1.15 $\mu$m and 1.10$-$1.31 $\mu$m.  Again, a 0$\farcs$38 slit was used to
obtain a resolving power of R$\approx$2500.  Conditions were 
0$\farcs$7$-$1$\farcs$0 seeing and clear skies.  Data
at both settings were acquired using two sets of dithered pairs with a
200-sec exposure at each dither position.  The shorter and longer
wavelength regions were acquired at airmasses of 1.34 and 1.57 for 2MASS
J2148+4003; the A0 dwarf calibrator HD 207220 was acquired in two nod
pairs having 4 coadds of 15s each and were taken at airmasses of
1.41 and 1.48 for the short-
and long-wavelength pieces, respectively.  Calibrations and reductions
were identical to those of the earlier NIRSPEC data described above, the
only difference being that the A0 star's hydrogen Paschen lines at 1.094
and 1.005 $\mu$m also needed to be interpolated over during the telluric
correction step.

\subsection{IRTF-SpeX Spectroscopy}

\subsubsection{Prism Mode: $\sim$0.7--2.5 $\mu$m}

Near-infrared spectra for 2MASS J1821+1414 and 2MASS J2148+4003 
were also obtained with SpeX 
\citep{rayner2003} on the 3.0-m NASA Infrared Telescope Facility.
2MASS J1821+1414 was observed in prism mode on 2005 Aug 10 UT with scattered 
clouds and $\sim$0$\farcs$7 seeing at $K$-band.  The prism mode provides 
continuous coverage from 0.7$-$2.5 $\mu$m in a single order on the
1024$\times$1024 InSb array.  Use of the $0{\farcs}5$ slit resulted in a 
resolving power of R${\approx}100$ in $J$-band
to $\approx$300 in $K$-band.  For
accurate sky subtraction, the target was observed in four  
nodded pairs having 120-second integrations per position at an airmass
of 1.03, for a total of 16 minutes of integration time.  The A0 star, HD
165029, was observed immediately prior to this observation at an
identical airmass followed by calibration flats and argon arc lamp exposures.

2MASS J2148+4003 was observed in prism mode on 2005 Sep 08 UT and 2005
Sep 09 UT with clear conditions and 0$\farcs$4 seeing at $H$-band on the
first night and heavy cirrus with 0$\farcs$5--0$\farcs$7 seeing at
$H$-band on the second night.  
On both nights, the target was observed in four nodded pairs having 
120-second integrations per position, for a total integration time of 16
minutes on each night.  The A0 star, HD 207220, 
was observed immediately 
after the science observations 
at a similar airmass ($\Delta$z~$<$~0.1) followed by
calibration flats and argon arc lamp exposures.
For comparison to the spectrum of 2MASS J2148+4003 (see $\S$4.2), 
2MASSW J22443167+2043433  (hereafter 2MASS J2244+2043) 
was observed in prism mode on 2005 Sep 09 
UT with the same instrumental set-up as the 2MASS J2148+4003 
observations.  The target was observed in six nodded pairs
having 120-second exposures per position, for a total integration time
of 24 minutes.  The same observation of HD 207220 that was used for the
2MASS J2148+4003 Sep 09 data was also used for telluric correction and flux 
calibration of 2MASS J2244+2043.  

Standard reductions were employed using the Spextool package version
3.2 \citep{cushing2004,vacca2003}.  The two fully reduced spectra of
2MASS J2148+4003 from 2005 Sep 08 UT and 2005 Sep 09 UT were combined into one
final spectrum by interpolating the wavelength scale of the spectrum
from 2005 Sep 09 UT onto the 
wavelength scale of the 2005 Sep 08 UT spectrum and then averaging 
the fluxes.  

\subsubsection{Short-Wavelength Cross-Dispersed Mode: $\sim$0.9--2.4 $\mu$m}

2MASS J1821+1414 and 2MASS J2148+4003 were observed in short-wavelength  
cross-dispersed mode to provide higher-resolution spectra over 0.9$-$2.4
$\mu$m in six orders.  Use of the $0{\farcs}5$ slit resulted in a 
resolving power of R${\approx}1200$.
2MASS J1821+1414 was observed on 2005 Aug 10 UT at an airmass of 1.1 
in six nodded pairs 
having 300-second integrations per position, 
for a total of 60 minutes of integration time.  
2MASS J2148+4003 was observed on 2005 Sep 09 UT at an airmass of 1.1 in
12 nodded pairs having 120-second integrations per position,
for a total of 48 minutes of integration time.

Observations of the A0 dwarf HD 165029 (for 2MASS J1821+1414) and the 
A0 dwarf HD 207220 (for 2MASS J2148+4003) near a similar
airmass and time as the target were taken with the above 
observations to provide both telluric correction and flux
calibration. For all observations the
instrument rotator was positioned at the parallactic angle. 
For instrumental calibration, internal flat-field and argon arc lamps 
were observed immediately after the observation of the A0 stars. 
Standard reductions were employed using the Spextool package version 
3.2. 

\subsubsection{Long-Wavelength Cross-Dispersed Mode: $\sim$1.9--4.2 $\mu$m}

Spectra of 2MASS J1821+1414 and 2MASS J2148+4003 were also obtained with SpeX 
using the long cross-dispersed mode, which provides coverage over 
1.9$-$4.2 $\mu$m in six orders.  
2MASS J2148+4003 was observed on 2006 Sep 01 UT, 2006 Sep 12 UT, and 
2007 Nov 11 UT.  The night of 2006
Sep 01 was clear with $0{\farcs}4$ seeing at $K$-band, the night of 
2006 Sep 12 was
mostly clear with light cirrus, high relative humidity ($\sim$75\%) and
$0{\farcs7}$ seeing at $K$-band, and the night of 2007 Nov 11 was clear with 
$0{\farcs9}$ seeing at $K$-band.  
Use of the $0{\farcs}5$ slit resulted in a resolving
power of R${\approx}1500$.  During the first
night, the target was observed in 20 on-slit nodded pairs having
30-second integrations per position, totalling 20 minutes of
integration; on the second night, the target was observed in three sets of
20 nodded pairs having 30-second integrations per position, totalling 60
minutes of integration; and on the third night, the target was observed in
five sets of 20 nodded pairs having 30-second integrations per position, 
totalling 100 minutes of integration.  For multiple sets on a single
night, the position angle of the slit was rotated to match the
parallactic angle between each set and between science and standard 
observations.  
For telluric correction and flux calibration,
the A0 dwarf HD 209932 was observed on the first two 
nights during a similar time 
and near the same airmass ($\Delta$z~$<$~0.1) 
as 2MASS J2148+4003.  During the third night
the A0 dwarfs HD 209932 and HD 13936 were observed at three different
times to match airmasses ($\Delta$z~$<$~0.15) of the five different sets of
science observations.  Immediately after the A0 observations,
we obtained flat-field and argon arc lamps for calibrations.  

2MASS J1821+1414 was observed in long cross-dispersed 
mode on 2007 Jul 28 UT.  The
night was clear with $0{\farcs8}$ seeing at $K$-band during observations.  
The target was observed in two sets of 20 nodded pairs having 30-second
integrations per position, totalling 40 minutes of integration.  The A0
dwarf, HD 165029, was observed after the 
science target followed immediately by calibration flats
and argon arc lamp observations.  
The airmass matches between the science and standard observations were 
very poor ($\Delta$z$\sim$0.5-0.7) since 2MASS J1821+1414 was observed until 
it was setting.  

Reductions were carried out using the Spextool package version 3.4.  
Due to the faint signal for both science
targets, we first combined the spectra of each nod position for all sets,
defined the apertures, wrote traces, extracted the spectra, and
then combined the AB aperture-stacked spectra.  The spectrum from 
each set was telluric
corrected using the standard closest in time and airmass.  
All six orders were then merged starting in $K$-band (order 10) and
going to progressively longer wavelengths (smaller orders) with no
autoscaling employed.   All sets were merged into one final
spectrum and regions of heavy telluric absorption, i.e., $\sim$2.5--3.0
$\mu$m, were removed.  Finally, we rebinned the spectra by a factor of
three such that one pixel corresponds to a single resolution element. 

\subsection{Spitzer-IRS Spectroscopy}

2MASS J2148+4003 and 2MASS J1821+1414 were observed (\textit{Spitzer} AORs  
16201984 and 16176128) using the Short-Low (SL) module of the  
Infrared Spectrograph (IRS; \citealt{houck2004}) onboard the \textit 
{Spitzer Space Telescope} \citep{werner2004}.  This module covers  
5.2$-$15.3 $\mu$m at R$\approx$ 90 in 2 orders.  
The observations consisted of a series of  
exposures taken at two different positions along each slit.  The  
total exposure time for each target was 3904 sec.  The raw data were  
processed with the IRS pipeline (version S13) at the \textit{Spitzer}  
Science Center and were further reduced as described in Cushing et  
al. (2006).  Briefly, the spectra were extracted with a fixed-width  
aperture of 6$''$ and observations of $\alpha$ Lac obtained as part  
of the IRS calibration observations were used to remove the  
instrument response function and to flux calibrate the observations of  
2MASS J2148+4003 and 2MASS J1821+1414.  

\section{Analysis}

\subsection{Red-Optical Spectral Classification}

\subsubsection{2MASS J1821+1414 (L4.5)}

\begin{figure}
\epsscale{0.7}
\plotone{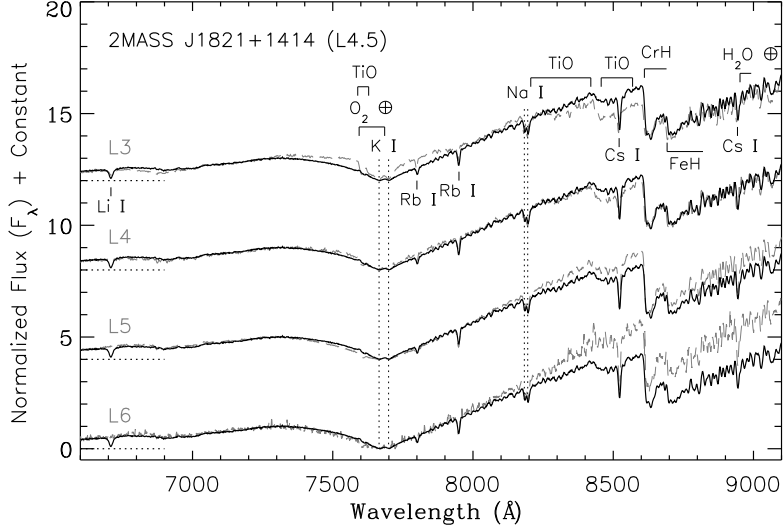}
\caption{Optical spectra from Subaru-FOCAS of 2MASS J18212815+1414010
(black, solid) compared to the typical field L dwarfs (gray, dashed): 
2MASSW J1146345+223053 (L3) and 2MASSW J1155009+230706 
(L4), taken with Keck-LRIS (R$\sim$1200),
and 2MASSW J1507476$-$162738 (L5) and 2MASS J21321145+1341584 
(L6), taken with Subaru-FOCAS (R$\sim$1000).  
The spectrum of 2MASS J1821+1414 has been telluric corrected while the spectra 
of the comparison L dwarfs have not.  All spectra have
been normalized to 1 using the median in a 20 \AA~window centered on
7300 \AA~and are separated along the vertical axis by a constant of 4.0
(dotted lines show zero levels) for clarity.  The continuum of 2MASS J1821+1414, 
shortward of 8600~\AA, appears to be intermediate between the L4 and L5.
\label{opt1821}}
\end{figure}

We spectroscopically classify 2MASS J1821+1414 in the 
red-optical ($\lambda<$~8600~\AA) 
by comparing its continuum shape in Figure~\ref{opt1821} 
to that of other known, optically classified field dwarfs: 
2MASSW J1146345+223053 (L3, hereafter 2MASS J1146+2230), 
2MASSW J1155009+230706 (L4, hereafter 2MASS J1155+2307), 
2MASSW J1507476$-$162738 (L5, hereafter 2MASS J1507$-$1627; \citealt{kirkpatrick1999}), 
and 2MASS J21321145+1341584 (L6, hereafter 2MASS J2132+1341; \citealt{cruz2007}).  
Spectra of 2MASS J1507$-$1627 and 2MASS J2132+1341, like 2MASS J1821+1414, were 
obtained using Subaru-FOCAS 
(R$\sim$1000), while spectra of 2MASS J1146+2230 and 2MASS J1155+2307 were obtained using Keck 
I-LRIS (R$\sim$1200).  
2MASS J1821+1414 has been telluric corrected while the comparison L dwarfs 
have not been telluric corrected.  The best continuum fit 
shortward of 8600 \AA~ 
appears to be intermediate between the L4 and L5 dwarfs, including 
the strengths of the TiO bandheads, 
as well as  the Rb I and Cs I atomic line strengths.  
We therefore adopt an optical spectral classification 
of L4.5 for 2MASS J1821+1414.  

Quite unlike the L dwarfs shown for comparison, 2MASS J1821+1414 shows deep 
Li absorption at 6708 \AA.  Using the IRAF package SPLOT, we 
measured an equivalent 
width (EW~=~13.9~$\pm$~0.4 \AA) of this feature by taking the 
average of five measurements and estimating 
the error by the standard deviation in these measurements added 
in quadrature to the 
average of five measurements of a typical noise spike.  
Approximately 40\% of L4--L4.5 dwarfs have a detected Li I absorption line
(EW$>$4 \AA) with a median EW of 7--9 \AA~(Kirkpatrick et al$.$, submitted).  
None of the L4--L4.5 dwarfs in that 
paper have Li EW measures as large as that seen in 2MASS J1821+1414.

\subsubsection{2MASS J2148+4003 (L6)}

\begin{figure}
\epsscale{0.7}
\plotone{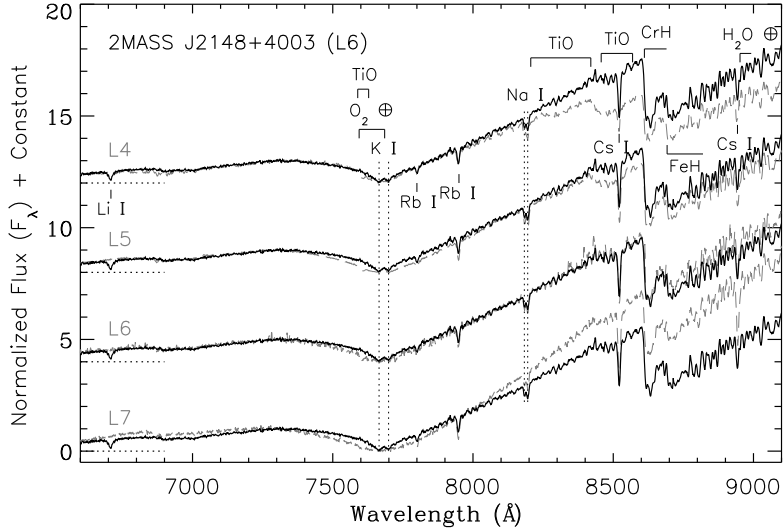}
\caption{Optical spectra from Subaru-FOCAS of 2MASS J21481628+4003593
(black, solid) compared to the typical field L dwarfs (gray, dashed): 
2MASSW J1155009+230706 (L4), taken with
Keck-LRIS (R$\sim$1200), 2MASSW J1507476$-$162738 (L5),
2MASS J21321145+1341584 (L6), and DENIS-P J0205.4$-$1159 (L7), 
taken with Subaru-FOCAS (R$\sim$1000).  
The spectrum of 2MASS J2148+4003 has been telluric corrected while the spectra 
of the comparison L dwarfs have not.  All spectra have been
normalized to 1 using the median in a 20 \AA~window centered on 7300
\AA~and are separated along the vertical axis by a constant of 4.0
(dotted lines show zero levels) for clarity.  The continuum of 2MASS
2148+4003, shortward of 8600 \AA, appears to match well to that of the L6.
\label{opt2148}}
\end{figure}

We determine the red-optical ($\lambda<$~8600 \AA) spectral type of 2MASS 
2148+4003 by comparing 
its continuum to that of known optically classified L dwarfs 
(Figure~\ref{opt2148}) -- 2MASS J1155+2307 (L4), 2MASS J1507$-$1627 (L5), 
2MASS J2132+1341 (L6), and DENIS-P J0205.4$-$1159 (L7, 
hereafter DENIS J0205$-$1159; \citealt{kirkpatrick1999}).  
DENIS J0205$-$1159 was also observed with Subaru-FOCAS 
using an identical set-up as that for 2MASS J2148+4003 but, like the other 
comparison L dwarfs, was not telluric-corrected whereas 2MASS J2148+4003 
was.  The continuum of 2MASS J2148+4003 shortward of 8600 \AA~matches 
very well to 
that of the L6 dwarf, including the TiO bandheads, while the atomic line 
strengths of Rb I, Na I, and Cs I are slightly weaker.  We therefore adopt an 
optical spectral type of L6 for 2MASS J2148+4003.

2MASS J2148+4003 also shows Li absorption in its spectrum. 
Using the same method described for 2MASS J1821+1414, we 
measure an EW of 12.1~$\pm$~0.6 \AA.  Nearly 70\% of L6--L6.5 dwarfs
have Li detected at greater than 4 \AA~EW in their spectra, and in these the 
EW ranges from 12--18 \AA~(Kirkpatrick et al., submitted).  
2MASS J2148+4003 appears to be a normal, Li-bearing L6.5. 

\subsection{Near-Infrared Spectral Classification}

\subsubsection{2MASS J1821+1414 (L5 pec)}

\begin{figure}
\epsscale{0.5}
\plotone{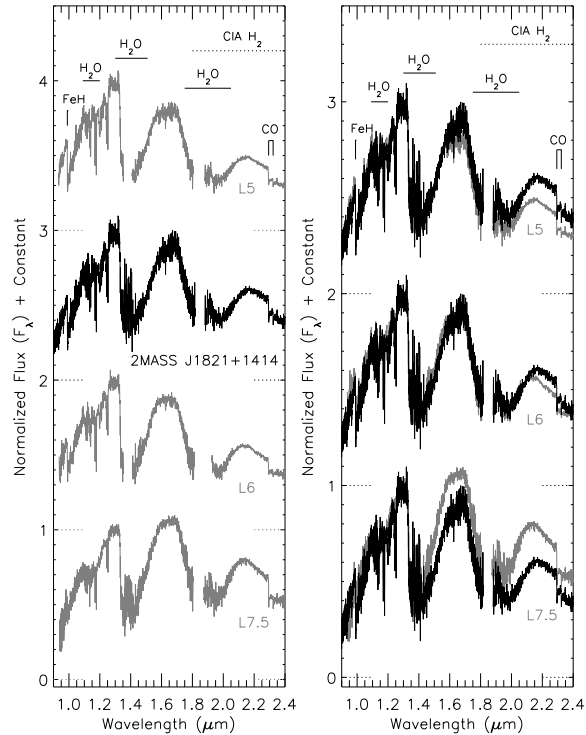}
\caption{{\it Left}: Near-infrared cross-dispersed spectra from 
IRTF-SpeX of 2MASS J18212815+1414010 
(black) plotted in comparison to normal, field L dwarfs (grey): 
2MASSW J1507476$-$162738 (L5), 2MASSW J1515008+484742 (L6), and 
2MASSI J0825196+211552 (L7.5) (from \citealt{cushing2005}).  
All spectra have been normalized to 1
using the median in a 0.004 $\mu$m window centered on 1.27 $\mu$m and are
separated along the vertical axis by a constant of 1.0 (dotted lines
show zero levels)
for clarity.  All spectra have comparable resolving powers, R$\sim$1200.  
{\it Right}: The same normalized spectra of field L dwarfs (gray) 
overlain with the spectrum of 2MASS J1821+1414 (black).  While the spectrum of
2MASS J1821+1414 is slightly redder and has a somewhat peakier $H$-band,
its gross spectral shape and H$_2$O absorption strengths fit closest to
the L6.
\label{NIR1821}}
\end{figure}

We compare the IRTF-SpeX cross-dispersed spectrum (R$\sim$1200) 
of 2MASS J1821+1414 to a spectral sequence of normal, late-type L field 
dwarfs from \cite{cushing2005} in Figure~\ref{NIR1821}.  
The spectral types of these late-type L dwarfs are from the optical
classification scheme \citep{kirkpatrick1999}.
The lowest spectral resolving power for this set is R$\approx$1200 for
2MASSI J0825196+211552 (L7.5, hereafter 2MASS J0825+2115; 
\citealt{kirkpatrick2000}) and for our object 2MASS J1821+1414.  
We have degraded the resolving power, R$\approx$2000, of 2MASS 
J1507$-$1627 and 2MASSW J1515008+484742 
(L6, hereafter 2MASS J1515+4847; \citealt{cruz2007}) 
to R$\approx$1200 by convolving the
spectra with a gaussian of appropriate FWHM to reach the desired
resolving power.  Comparable spectral resolving powers are important
when visually comparing very
narrow features such as atomic lines, which change depth according to
the resolving power used.
  
Examining Figure~\ref{NIR1821}, we see that the overall spectral 
morphology of 2MASS J1821+1414 falls intermediate between the L6 
and L7.5 but is most similar to that 
of the L6, having comparable flux levels at $J$- and $H$-bands and similar
H$_2$O absorption strengths.  Nonetheless, there are notable differences
between it and the L6 standard $-$ the triangular-shaped $H$- and $K$-band 
continua, the stronger CO bands near 2.3 $\mu$m, the slightly shifted
$K$-band peak, and the overall redder spectrum.  

\begin{figure}
\epsscale{0.7}
\plotone{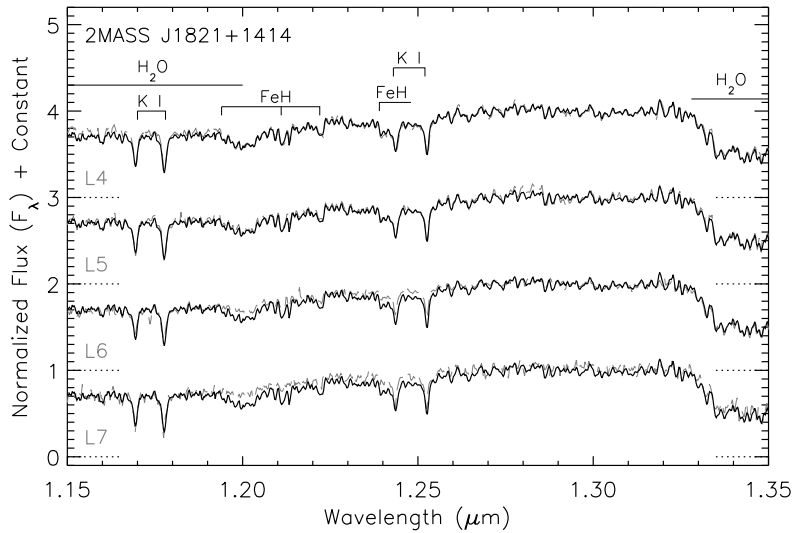}
\caption{1.14--1.38 $\mu$m Keck-NIRSPEC data (R$\sim$2000) of
2MASS J18212815+1414010 (black, solid) overlaid on a grid of mid-to
late-type L dwarfs (gray, dashed): 2MASS
J21580457$-$1550098 (L4), DENIS-P J1228.2$-$1547 (L5), 
2MASSs J0850359+105716AB (L6), and 2MASSW J1728114+394859 (L7), taken with
the same telescope/instrument set-up 
\citep{mclean2003}.  All spectra have been normalized to 1 at 1.27
$\mu$m using the median in a 0.004 $\mu$m window.  The spectra are separated
vertically in integer units of 1, with zero levels shown by dotted lines.  
The continuum of 2MASS J1821+1414 in 
this window and the K I and FeH absorption strengths match closest to the L5.  
\label{zoom1821}}
\end{figure}

Since it is not uncommon for near-infrared spectral types to differ from
optically derived spectral types for the same object (see section 3.2 of
\citealt{kirkpatrick2005}), we now focus on the $J$-band portion of the  
spectra of 2MASS J1821+1414 in Figure~\ref{zoom1821}.  We compare
the Keck-NIRSPEC spectrum of 2MASS J1821+1414 to a grid of optically classified 
L dwarfs acquired by \cite{mclean2003} with the same telescope/instrument 
setup: 2MASS J21580457$-$1550098 (L4), DENIS-P J1228.2$-$1547 (L5, hereafter 
DENIS J1228$-$1547), 2MASSs J0850359+105716AB (L6, hereafter 2MASS J0850+1057AB), 
and 2MASSW J1728114+394859 (L7, hereafter 2MASS J1728+3948).  
Over this wavelength regime, the continuum from
1.26--1.32 $\mu$m, the strength of the H$_2$O-band redward of 1.32, and
the strength of the FeH bands most closely match the L5.  In fact, the
agreement over this region is far better than that seen for any of the
standards in the $H$- or $K$-band windows.   

We adopt an overall near-infrared spectral type of L5 pec for 2MASS
J1821+1414 based on its spectral characteristics at $J$-band and
peculiarities at longer wavelengths.  We note
that the spectrum itself deviates from a standard L5 as the
$H$-band portion is markedly more peaked than in the standard and the
overall spectrum is slightly redder.  We discuss
possible causes for these differences in later sections.

\subsubsection{2MASS J2148+4003 (L6.5 pec)}

\begin{figure}
\epsscale{0.5}
\plotone{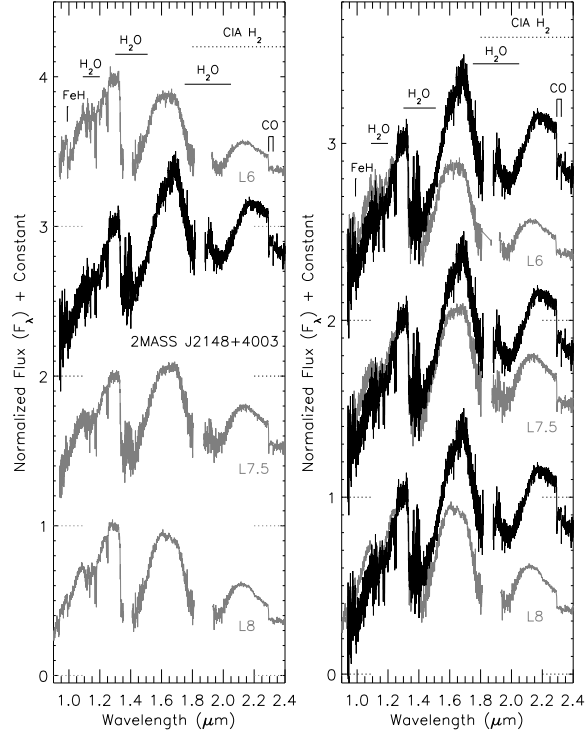}
\caption{{\it Left}: IRTF-SpeX cross-dispersed spectra of 2MASS
J21481628+4003593 (black) plotted in comparison to normal field L dwarfs
(gray): 2MASSW 
J1515008+484742 (L6), 2MASSI J0825196+211552 (L7.5), and 
DENIS-P J0255$-$4700 (L8) (from 
\citealt{cushing2005}).  All spectra have been normalized to 1
using the median in a 0.004 $\mu$m window centered on 1.27 $\mu$m and are
separated along the vertical axis by a constant of 1.0 (zero levels are
shown by dotted lines) for clarity.  
All spectra have comparable resolving powers, R$\sim$1200.  {\it Right}: 
The same normalized spectra of field L dwarfs (gray) overlain with the spectrum 
of 2MASS J2148+4003 (black).  The spectrum of 2MASS J2148+4003 is clearly
divergent from the late-type L dwarfs, with a much redder overall
spectrum; redder overall slope in the 0.9--1.3 $\mu$m region; 
triangular shaped $H$- and $K$-band continua; stronger CO band absorption at
2.29 and 2.32 $\mu$m; and a red-shifted $K$-band peak.
\label{NIR2148}}
\end{figure}

We follow a similar prescription of spectral analysis for 2MASS
J2148+4003 as done above for 2MASS J1821+1414.  In Figure~\ref{NIR2148},  
we compare the IRTF-SpeX cross-dispersed spectrum (R$\sim$1200) 
of 2MASS J2148+4003 to a spectral sequence of normal, late-type L field
dwarfs from \cite{cushing2005} -- 2MASS J1515+4847 (L6),   
2MASS J0825+2115 (L7.5), and DENIS-P J0255$-$4700 (L8, hereafter DENIS J0255$-$4700; 
\citealt{martin1999}).  We degrade the resolving power, R$\sim$2000, of both 2MASS 
1515 and DENIS J0255$-$4700 to match the resolving power, R$\sim$1200,
of 2MASS J0825+2115 (L7.5) and 2MASS
J2148+4003.  Figure~\ref{NIR2148} highlights the extreme peculiarities of 2MASS
J2148+4003 -- the marked redness of the NIR spectrum, the
triangular-shaped $H$- and $K$-band continua, an unusually red slope throughout the 
$Z$-, $Y$-, and $J$-bands, the red-shifted $K$-band peak and the large CO-band
features at 2.29 \& 2.32 $\mu$m.  These two panels show that 2MASS
J2148+4003 is unlike any of these late-type L dwarfs.

\begin{figure}
\epsscale{0.7}
\plotone{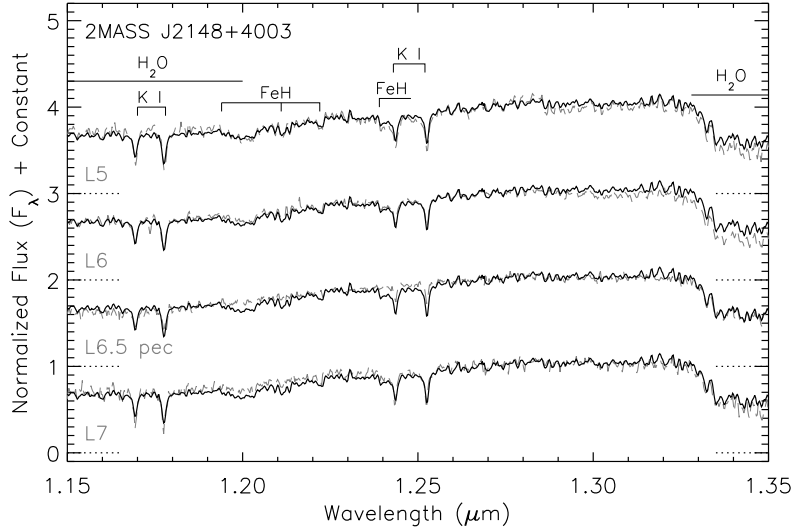}
\caption{1.14--1.38 $\mu$m Keck-NIRSPEC data (R$\sim$2000) of
2MASS J21481628+4003593 (black, solid) overlaid on a grid of 
late-type L dwarfs (gray, dashed): DENIS-P
J1228.2$-$1547 (L5), 2MASSs J0850359+105716AB (L6), 2MASSW J1728114+394859 
(L7) and the very red L6.5 pec 2MASSW J2244316+204343, taken with the
same telescope/instrument set-up \citep{mclean2003}.  All spectra have 
been normalized to 1 at 1.27 $\mu$m using the median in a 0.004 $\mu$m 
window.  The spectra are separated vertically in integer units of 1,
with zero levels  shown by dotted lines.  The continuum shape of 
2MASS J2148+4003 is well matched to that of 2MASS J2244+2043 but diverges from it 
in the absorption strengths of K I and FeH.
\label{zoom2148}}
\end{figure}

Now we turn to the $J$-band spectrum in Figure~\ref{zoom2148}, which
compares the Keck-NIRSPEC spectrum of 2MASS J2148+4003 with late-type 
optically classified L dwarfs taken with the same telescope/instrument 
set-up \citep{mclean2003}: DENIS J1228$-$1547 (L5), 2MASS J0850+1057AB (L6), 
2MASS J+39481728 (L7) and the very red L6.5 pec 2MASS J2244+2043.  
The continuum shortward of 1.26
$\mu$m and the K I \& FeH absorption strengths match the L6
closest, although it should be noted that the L6 used here, 
2MASSI J0103320+193536 \citep{kirkpatrick2000}, 
has weaker K I lines and FeH band strengths than
expected in the late-type L sequence.  From 1.26 to 1.32 $\mu$m, the
continuum of 2MASS J2148+4003 more closely matches the L7.  Throughout
this window, the continuum of 2MASS J2148+4003 and the H$_2$O band 
strengths match very well to the L6.5 (NIR L7.5 pec) dwarf 
2MASSW J2244316+204343 (hereafter 2MASS J2244+2043; \citealt{mclean2003}).
However, the two spectra do deviate in their
finer structure -- 2MASS J2244+2043 has very weak K I lines and 
indiscernible FeH absorption bands, presumably because it is a
low-gravity object \citep{mclean2003}.  

Using the results of our fits
from the $J$-band windows and the overall peculiarities at $H$- and  
$K$-bands, we adopt an overall spectral type of
L6.5 pec for 2MASS J2148+4003.  We discuss
possible causes for the peculiarities in $\S$4.

\subsection{5.2--15.3 $\mu$m Spectroscopic Features}

\begin{figure}
\epsscale{0.7}
\plotone{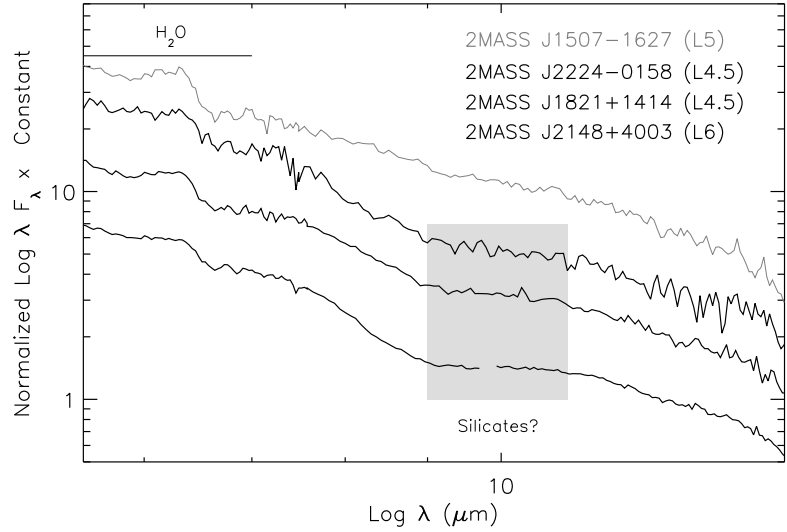}
\caption{Spitzer-IRS spectra (R$\sim$90) of 2MASSW J1507476$-$162738 
(optical L5; gray, top), 2MASSW J2224438$-$015852 (optical L4.5; black,
second from top), 
2MASS J18212815+1414010 (optical L4.5; black, second from bottom), and 
2MASS J21481628+4003593 (optical L6; black, bottom).  All spectra have been normalized 
to 1 at 6 $\mu$m using the median in a 0.4 $\mu$m window.  The spectra have 
been offset by multiplicative constants of 2.  2MASS J1507$-$1627 is a normal L5 dwarf, 
while 2MASS J2224$-$0158, 2MASS J1821+1414, and 2MASS J2148+4003 are redder than typical L dwarfs 
of similar spectral type.  These three objects show a flattening at 9--11 $\mu$m, 
which \cite{cushing2006} have tentatively identified as silicate absorption.
\label{IRSspectra}}
\end{figure}

Figure~\ref{IRSspectra} shows the IRS spectrum of 2MASS J1821+1414
and 2MASS J2148+4003 
along with the spectrum of 2MASS J1507$-$1627 (L5) and 2MASSW J2224438$-$015852
(L4.5, hereafter 2MASS J2224$-$0158; \citealt{kirkpatrick2000}) from  
\cite{cushing2006}.  The spectra of 2MASS J2148+4003, 2MASS J1821+1414, and  
2MASS J2224$-$0158 all exhibit a broad absorption feature from 9--11 $\mu$m  
that is absent (or rather, weak) in the spectrum of 2MASS J1507$-$1627.   
\cite{cushing2006} tentatively identified this feature in the  
spectrum of 2MASS J2224$-$0158 as arising from the Si-O stretching mode of  
the silicate grains.  
\cite{burgasser2008} have also noted that 2MASS J2224$-$0158 (L4.5) is  
$\sim$0.5 mag redder in $J-K_s$ than 2MASS J1507$-$1627 (L5), perhaps indicating a  
correlation between the thickness of the condensate clouds and the  
NIR colors of L dwarfs.  Both 2MASS J2148+4003 and 2MASS J1821+1414 are $\sim$0.1 and 
$\sim$0.5 mags redder than the median $J-K_s$ values of L4--L4.5 and L6--L6.5 
dwarfs, respectively, which appears to strengthen this conclusion.

\subsection{Spectro-Photometric Distance Estimates}

When estimating distances, optical spectral types are preferred over NIR 
spectral types because they 
are a better predictor of M$_J$, particularly for mid-to late-type L dwarfs 
(see Figure 9 of \citealt{kirkpatrick2005}).  We use the M$_J$ versus 
spectral type relation derived by \cite{looper2008} to 
estimate the distance to 2MASS J1821+1414 and
2MASS J2148+4003. Using the L4.5 optical spectral type and the 
2MASS-measured value of $J$=13.43$\pm$0.03 mag for 2MASS J1821+1414, we estimate 
a distance of 9.8~$\pm$~1.3 pc if it is single. The optical spectral type
of L6 along with the 2MASS-measured value of $J$=14.15$\pm$0.04 mag for 2MASS J2148+4003
suggest a distance of 9.9~$\pm$~1.3 pc if it is single.  Trigonometric parallax 
measurements are needed to confirm these estimates.  

To place these discoveries in context we have listed them in Table 3
along with the eight other L dwarfs known or believed to lie
within 10 parsecs.  Of these eight 
L dwarfs, only three have trigonometric parallax measurements.  
For the five 
objects which do not have trigonometric parallaxes, we use the 
\cite{looper2008} relation to spectrophotometrically 
estimate their distances and list these in Table 3, as well.  
As with 2MASS J1821+1414 and 2MASS J2148+4003, trigonometric parallaxes are needed to 
confirm these distance estimates.

The discovery of these two objects has increased the number of L dwarfs
within 10 pc by $\sim$25\%, a substantial increase in this nearest
census of the Solar Neighborhood.  However, we are not taking into
account Malmquist bias when discussing the number of objects within 10
pc.  These two discoveries have arisen because
of the sky coverage of our proper motion survey extending into a
relatively unexplored part of the Galactic Plane
($10^\circ \le |b| \le 15^\circ$).

\section{Discussion}

\subsection{Dusty Photospheres}

All available evidence indicates that 2MASS J1821+1414 and 2MASS J2148+4003 have 
unusually dusty photospheres -- their unusual red slopes throughout the 
$Z$-, $Y$-, and $J$-bands, overall red NIR  
colors, weak H$_2$O absorption, and the presence of 
silicate absorption at 9--11 $\mu$m.
The wavelength dependence of dust absorption and scattering (i.e., dust 
extinction) causes an object to appear redder than expected, because 
shorter wavelength light is attenuated and scattered more than longer 
wavelength light.  
For absorption bands that form in atmospheric layers near or below  
the cloud deck (e.g., the H$_2$O absorption bands at $Y$- and $J$-band), the  
additional opacity provided by the condensates limits the depth in  
the atmosphere from which radiation can escape and thus weakens  
their relative strengths.  The atmospheric layers above the cloud  
decks are also warmed due to the presence of the condensate clouds.  
Absorption lines (e.g., K I) and bands (e.g., H$_2$O) 
that form in these atmospheric layers  
are therefore also weakened.  Lastly, the flattening seen from 
9--11 $\mu$m in the spectra of 2MASS J1821+1414 and 
2MASS J2148+4003 is likely due 
to a higher bulk content of silicate dust grains in the atmosphere or 
from smaller grain sizes than are typical for field L dwarfs.  In the
next section, we suggest two underlying physical causes for dusty photospheres.  

\subsection{The Nature of Both Objects}

Below, we offer four possible explanations for the peculiar spectral
features of 2MASS J1821+1414 and 2MASS J2148+4003, two of which may
lead to dusty photospheres:

{\bf (1) Unresolved Binarity:} In the case of 2MASS J2148+4003, we
    can eliminate the possibility of unresolved binarity being the
    underlying cause for its extreme red color ($J-K_s$=2.38).  The
    composite $J-K_s$ color of a binary results in a color intermediate
    between the two components and so the composite cannot be redder
    than either individual member.  Since 2MASS J2148+4003 is one of only 
    seven L or T dwarfs with $J-K_s>$~2.3, out of 648 objects\footnote{See 
    http://dwarfarchives.org.}, two ordinary L or T dwarf components
    would necessarily be bluer and could not combine to produce a redder
    composite. 

    Also, no known combination of L+L or L+T dwarfs can reproduce the
    triangular-shaped $H$-band present in the spectra of 2MASS
    J1821+1414 
    and 2MASS J2148+4003 (See Figure 4 of \citealt{burgasser2007b}), 
    so we must account for these features in some other way.

{\bf (2) Interstellar Reddening:} Interstellar material, because of its
    wavelength-dependent extinction properties, can cause the spectral
    energy distribution of a background star seen through it to be
    artificially reddened.  However, the proper motions of these two objects
    alone imply that they are nearby and well within the ``Local Bubble'' 
    \citep{snowden1998}; hence, there is no significant amount of
    interstellar material expected to be between us and these dwarfs.

{\bf (3) Unusual Metallicity:} Is it possible that these two objects  
    represent unusually low- or high-metallicity L dwarfs? Several low  
    metallicity L dwarfs have already been catalogued, the most relevant  
    examples being the sdL4 2MASS J16262034+3925190 (hereafter 2MASS J1626+3925; 
    \citealt{burgasser2004}) and the sdL7 2MASS J05325346+8246465 
    (hereafter 2MASS J0532+8246; \citealt{burgasser2003}). In both  
    cases these objects are much bluer, not redder, than field L dwarfs  
    of the same type, the 2MASS $J-K_s$ values being $-$0.03$\pm$0.08  
    mag and 0.26$\pm$0.16 mag for 2MASS J1626+3925 and 2MASS J0532+8246, 
    respectively. Moreover, the near-infrared spectra of these  
    objects do not show spectra that are sharply peaked at the 
    long-wavelength side of $H$-band, as seen in 2MASS J1821+1414 
    and 2MASS J2148+4003, but instead show $H$-band spectra that are 
    very rounded. 

\begin{figure}
\epsscale{0.8}
\plotone{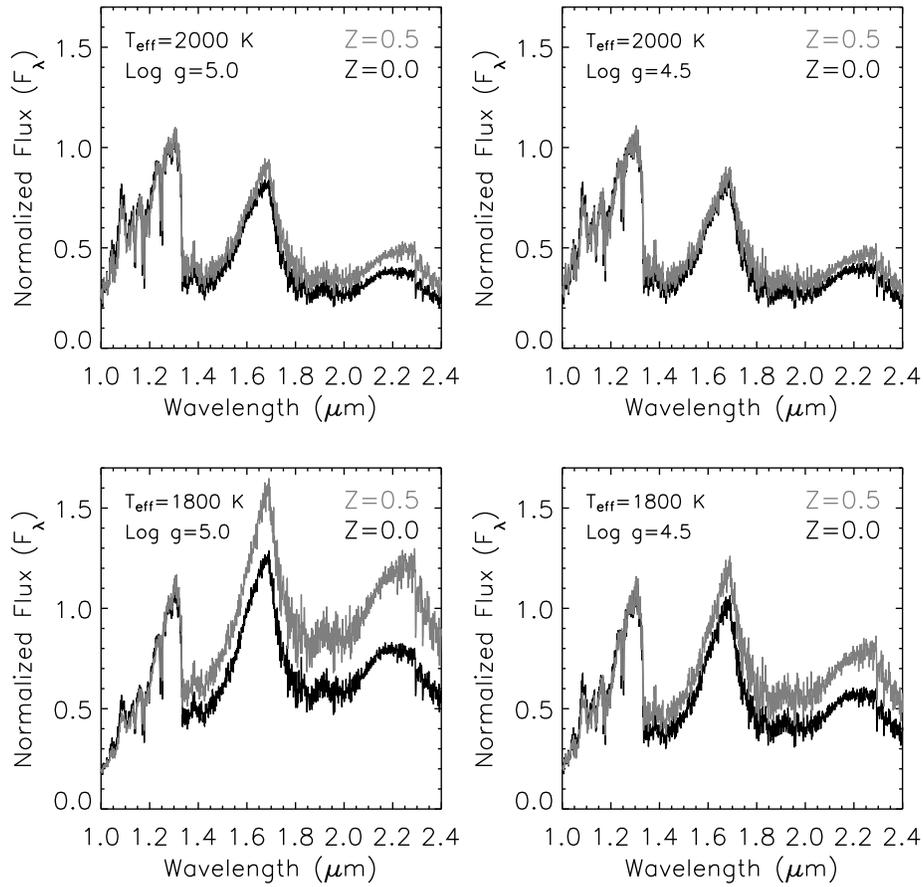}
\caption{Comparison of GAIA models (R$\sim$1200), showing differences between 
solar metallicity Z~=~[Fe/H]~=~0.0 (black) and metal-rich Z~=~0.5 (gray) models 
at fixed effective temperatures and gravities.  Note that the metal-rich models 
have comparatively redder spectra, weaker atomic lines, and weaker H$_2$O 
absorption.
\label{zmodels}}
\end{figure}

    High metallicity may imply greater dust production since more metals 
    are available to condense onto grains.  
    We examine GAIA models of high metallicity Z~=~[Fe/H]~=~0.5 
    versus solar metallicity Z~=~0.0 in 
    Figure~\ref{zmodels}.  The cases shown in this figure are for 
    effective temperatures and surface gravities typical for mid-type L dwarfs.  
    At a fixed T$_{eff}$ and surface gravity, the metal-rich models have 
    overall redder spectra, weaker FeH absorption, weaker Na I and K I 
    absorption, weaker H$_2$O absorption, and a red-shifted $K$-band peak   
    compared to the solar-metallicity models.  These are the same differences 
    noted between the two L dwarfs reported here, 2MASS J1821+1414 and 2MASS J2148+4003, 
    and typical mid-to late-type L dwarfs.  It is possible that these two 
    objects could be metal-rich, causing the spectroscopic peculiarities noted 
    here.  

{\bf (4) Unusual Gravity:} High gravity is not  
    favored because it gives rise to near-infrared spectra that are bluer  
    than normal and have spectral peaks in $H$- and $K$-bands that are rounded 
    in appearance due to increased collision induced absorption by H$_2$
    (CIA H$_2$).  CIA H$_2$ dominates the absorption at $H$- and
    $K$-bands \citep{borysow1997} and increases in strength with higher 
    pressures.  
    
    Low gravity, however, creates just the opposite effect. As explained in   
    \cite{kirkpatrick2006}, low gravity  
    results in lower atmospheric pressures, diminishing the relative  
    importance of CIA H$_2$.   With less competing opacity at $H$-band and  
    especially at $K$-band, flux can more easily emerge at these  
    wavelengths, making such low-gravity objects redder in NIR  
    ($JHK$) colors. Lower gravity may also tend to retard the precipitation  
    of condensates from the atmosphere, leading to thicker clouds, more
    condensate opacity, and redder NIR colors.  
    Reduced CIA H$_2$ and/or increased condensate opacity  
    are likely responsible for the triangular-shaped $H$-band spectra 
    characteristic of low-gravity, late-type M and early-type L dwarfs 
    seen both in young clusters (e.g., Orion: \citealt{lucas2001}; 
    Cha II, Rho Oph: \citealt{allers2007}) 
    and associations (e.g., \citealt{kirkpatrick2006}; 
    TWA: \citealt{looper2007b}). 
    
\begin{figure}
\epsscale{1.0}
\plotone{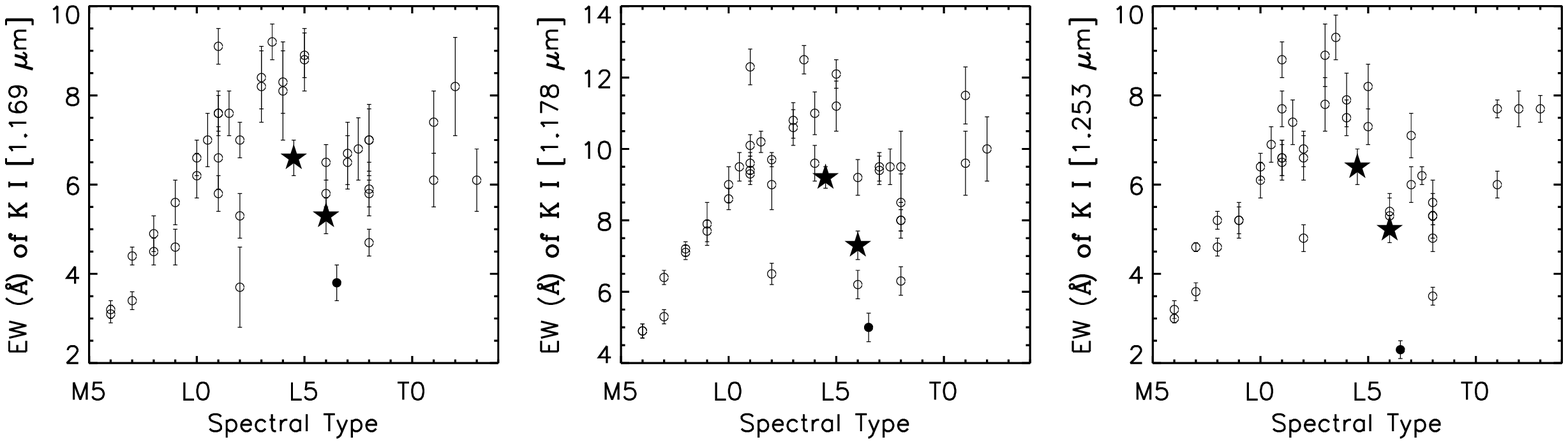}
\caption{Comparison of equivalent widths for the atomic line, K I, at
each indicated wavelength for objects with spectral types M6--T3.  
Field dwarfs from \cite{mclean2003} are
shown as open circles; 2MASS J2244+2043, also from \cite{mclean2003}, is shown
as a filled circle; and 2MASS J1821+1414 and 2MASS J2148+4003 are shown as filled
stars, with spectral types L4.5 and L6, respectively.  All data is taken
with Keck-NIRSPEC with a similar set-up.  Error bars are indicated.
\label{K1lines}}
\end{figure}

\begin{figure}
\epsscale{0.4}
\plotone{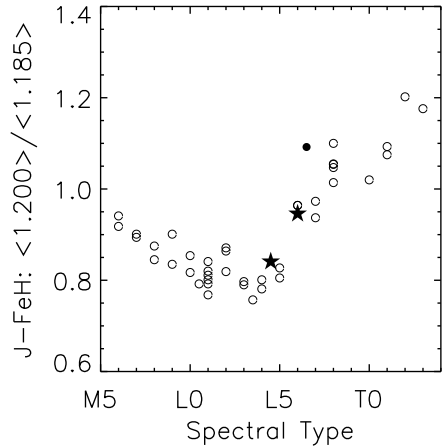}
\caption{Comparison of FeH strengths using the flux ratios over a 0.004 
$\mu$m window centered on the FeH absorption band at 1.200 $\mu$m and 
the pseudo-continuum at 1.185 $\mu$m, defined as the $J$-FeH flux ratio 
by \cite{mclean2003}.  The larger the flux ratio, the weaker the absorption.  
Field dwarfs from \cite{mclean2003} are
shown as open circles; 2MASS J2244+2043, also from \cite{mclean2003}, is shown
as a filled circle; and 2MASS J1821+1414 and 2MASS J2148+4003 are shown as filled
stars, with spectral types L4.5 and L6, respectively.  All data is taken
with Keck-NIRSPEC with a similar set-up.  
\label{FeHlines}}
\end{figure}

    Both 2MASS J1821+1414 and 2MASS J2148+4003 have $H$-band spectral 
    morphologies 
    similar to low-gravity objects.  Other gravity diagnostics such as 
    atomic K I line strengths and FeH absorption \citep{mcgovern2004} 
    may support this hypothesis.  In Figure~\ref{K1lines}, we plot the 
    line strengths of K I at 1.169, 1.178, \& 1.253 $\mu$m for 2MASS 
    J1821+1414 and 2MASS J2148+4003 in comparison to other L dwarfs, 
    using our Keck-NIRSPEC  
    data.  The K I line strengths of 2MASS J1821+1414 and 2MASS
    J2148+4003 are weaker 
    than those of other L dwarfs of similar spectral type but are not as extreme 
    outliers as the possibly low-gravity L dwarf 2MASS J2244+2043.  
    Similarly, the $J$-FeH flux ratio, defined by 
    \cite{mclean2003}, show somewhat weaker FeH absorption in 2MASS J1821+1414 
    and comparable FeH absorption in 2MASS J2148+4003 as in normal L dwarfs of 
    similar spectral type (see Figure~\ref{FeHlines}).  
    2MASS J2244+2043, on the other hand, shows markedly weaker FeH absorption.  
    The weaker atomic lines seen in the spectra of 2MASS J1821+1414 and
    2MASS J2148+4003 could also be a 
    consequence of the higher opacity caused by thick dust clouds in their 
    atmospheres, as explained in $\S$4.1.

    Based on their spectrophotometric distance estimates of 10 pc, we calculate
    the tangential velocities for both objects (see Table 4).  2MASS
    J1821+1414 has a low
    V$_\mathrm{tan}$ of $\sim$10 km s$^{-1}$, suggesting that it could
    indeed be young.  However, 2MASS J2148+4003 has a large 
    V$_\mathrm{tan}$ of $\sim$62 km s$^{-1}$ $-$ well above the median
    value for L dwarfs of 24.5 km s$^{-1}$ \citep{vrba2004}.  This large 
    tangential velocity suggests that 2MASS J2148+4003 is not young.

\subsection{Bolometric Luminosity Measures}

\begin{figure}
\epsscale{0.7}
\plotone{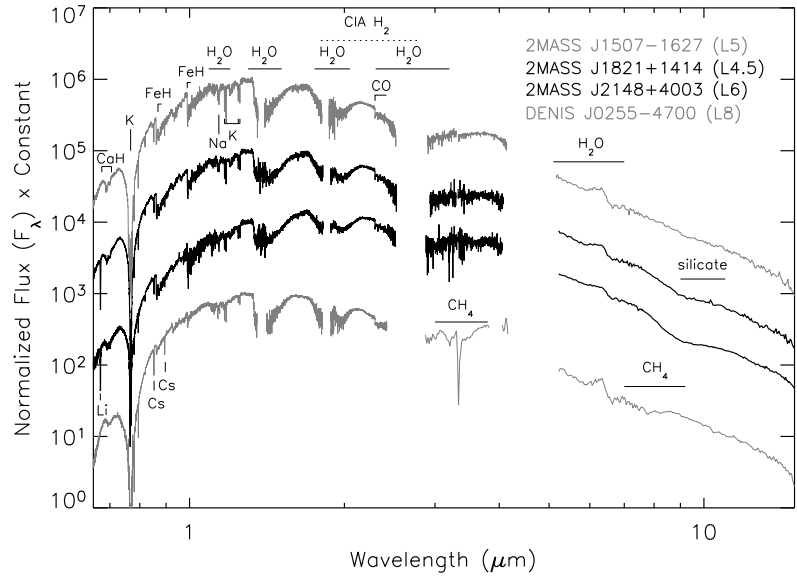}
\caption{The combined optical (Subaru-FOCAS), near-infrared (IRTF-SpeX), and 
mid-infrared (Spitzer-IRS) spectra of 2MASSW J1507476$-$162738 (optical
L5; gray, top), 2MASS J18212815+1414010 (optical L4.5; black, second
from top), 2MASS J21481628+4003593 (optical L6; black, second from bottom), 
and DENIS-P J0255$-$4700 (optical L8; gray, bottom).  
All spectra have been normalized to 1.0 using the median in a 0.04
$\mu$m window centered on 1.27 $\mu$m and are multiplied by constants to
clearly separate the spectra.  Important molecular and atomic features
are labeled throughout this regime.  
\label{SEDs}}
\end{figure}

In Figure~\ref{SEDs}, we show the nearly complete 0.6--15.3 $\mu$m 
spectroscopy of 2MASS J1821+1414 and 2MASS J2148+4003 in comparison 
to a typical L5 and L8 dwarf.  To create a
spectral energy distribution (SED), we interpolated between the
Subaru-FOCAS, IRTF-SpeX prism, IRTF-SpeX long cross-dispersed, and 
Spitzer-IRS spectra.  All data but the Spitzer/IRS spectrum were flux
calibrated with appropriate photometric measurements before being
stitched together.  No photometric measurements of 2MASS J1821+1414 
and 2MASS J2148+4003 in the 5.2--15.3 $\mu$m range were taken, but  
\cite{cushing2006} have shown that the absolute flux calibrations are good to 
$\sim$3\%.  Since no data was available
shortward of $\sim$0.6 $\mu$m, we interpolated between the flux level at
0.6 $\mu$m and a flux level of zero at 0 $\mu$m.  Longwards of 14.5
$\mu$m, we connected a Raleigh-Jeans tail.  The bolometric luminosity
was then computed by integrating underneath this SED and assuming 
distances of 9.8~$\pm$~1.3 pc and 9.9~$\pm$~1.3 pc for 2MASS J1821+1414 and
2MASS J2148+4003, respectively (see Table 4).

Note that, although 2MASS J2148+4003 has a nearly identical distance estimate as 
2MASS J1821+1414 ($\sim$10 pc) and has a later spectral type, its apparent 
bolometric magnitude is \textit{brighter} than 2MASS J1821+1414,
indicating that it 
is actually closer than 2MASS J1821+1414.  This is not surprising if this object 
does have a much dustier atmosphere, causing an abnormally faint $J$-band flux 
compared to typical L6 dwarfs.  Since the spectrophotometric distance estimate 
we used is based on $J$-band, this would overestimate the distance to
this object.  
In comparison, the M$_{K_s}$ versus spectral type relation derived by 
\cite{looper2008}, yields an identical distance estimate of 9.8~$\pm$~1.5 pc 
for 2MASS J1821+1414 but a closer distance estimate of 7.9~$\pm$~1.2 pc for 
2MASS J2148+4003.

\section{Conclusions}

We have presented spectral classification and distance estimates for two 
nearby and peculiar L dwarfs identified in our $\sim$4700 sq.\ degree 
2MASS proper motion search.  
With optical spectral types of L4.5 and L6, we estimate distances of 
$\sim$10 pc for both 2MASS J1821+1414 and 2MASS J2148+4003, although 
2MASS J2148+4003 may actually be closer.    
Both objects have overall red spectral energy distributions, 
weak H$_2$O bands, slightly triangular-shaped $H$-band continua, and 
silicate absorption at 9--11 $\mu$m.   
These features are likely due to unusually thick dust clouds in their 
atmospheres.  While the underlying cause for this dust is uncertain, we 
suggest that it may arise from a metal-rich atmosphere 
or a lower surface gravity, the latter implying a young age.  
The former cause is more likely in the case of 2MASS J2148+4003, 
whose large v$_{tan}$ indicates that it is not a young source.
Further follow-up is needed to distinguish between these two scenarios 
and to identify clear diagnostics for characterizing the detailed 
physical properties of other L dwarfs in the vicinity of the Sun.

\acknowledgements

We thank our referee, Nigel Hambly, for a timely and helpful report.  
We would like to thank our Keck observing assistants Steven Magee and
Julie Rivera and 
our IRTF observing assistants Paul Sears and Dave Griep for providing expert 
operation of the telescopes during our runs.  We thank 
our instrument scientists Jim Lyke and Grant Hill at Keck and Bobby Bus and 
John Rayner at IRTF for their expertise in running the spectrographs, 
imagers, and reduction software packages.  
We thank Takashi Hattori for providing instrument support at Subaru.  
We would like to thank John Stauffer and Maria Morales-Calderon 
for giving us some telescope time in exchange for instrument 
expertise on 2005 Dec 09 (UT) at Keck-II.  
DLL thanks John Rayner for advising her for part of this project, and
Dave Sanders for financial support.  
We are grateful to Lee Rottler for assistance with
setting-up the REDSPEC package at IPAC.  This paper uses data 
from the Brown Dwarf Spectroscopic Survey Archive 
(\url{http://www.astro.ucla.edu/$\sim$mclean/BDSSarchive}), 
the IRTF Spectral Library (\url{http://irtfweb.ifa.hawaii.
edu/$\sim$spex/spexlibrary/IRTFlibrary.html}), the SpeX Prism Library 
(\url{http://web.mit.edu/ajb/www/brown
dwarfs/spexprism/index.html}), and \url{http://DwarfArchives.org}.  
DLL would like to thank the
Northrop Grumman Corporation for a small grant to IPAC which paid for 
partial salary support during the search phase of this project. This 
publication also makes use of data products from the Two Micron All Sky 
Survey (2MASS), which is a joint project of the University of
Massachusetts and the Infrared 
Processing and Analysis Center/California Institute of Technology, 
funded by the National Aeronautics and Space Administration and the 
National Science Foundation. Guest User, Canadian Astronomy Data Centre, 
which is operated by the Herzberg Institute of Astrophysics, National 
Research Council of Canada.  This research has also made use of the 
NASA/IPAC Infrared Science Archive (IRSA), which is operated by the 
Jet Propulsion Laboratory, California Institute of Technology, under 
contract with the National Aeronautics and Space Administration. 
The 04 Aug 2006 UT NIRSPEC data were obtained at the W.M. Keck
Observatory from telescope time allocated to the National Aeronautics
and Space Administration through the agency's scientific parntership
with the California Institute of Technology and the University of
California.  The Observatory was made possible by the generous financial
support of the W.M. Keck Foundation.  As all 
spectroscopic and imaging follow-up data were obtained from the summit 
of Mauna Kea, the authors wish to recognize and acknowledge the very 
significant cultural role and reverence that this mountaintop has always 
had within the indigenous Hawaiian community. We are most fortunate to 
have the opportunity to conduct observations there.

\clearpage

\begin{deluxetable}{cccccc} 
\tabletypesize{\scriptsize}
\tablecolumns{6} 
\tablewidth{0pc} 
\tablecaption{2MASS Measurements for the Two New L Dwarfs}
\tablehead{ 
\colhead{Obs Date} & \colhead{RA(J2000)} & \colhead{Dec(J2000)} &
\colhead{$J$}   & \colhead{$H$}    & \colhead{$K_s$} \\
\colhead{(UT)} & \colhead{(deg)} & \colhead{(deg)} &
\colhead{(mag)} & \colhead{(mag)} & \colhead{(mag)}}
\startdata
\cutinhead{2MASS J18212815+1414010,
$\mu=0{\farcs}21\pm0{\farcs}$07 yr$^{-1}$ at
$\theta$=153.4$^{\circ}\pm$0.7$^{\circ}$, $\Delta$t=1.13 yr}
1999 Mar 27\tablenotemark{a} & 275.367305 & +14.233623 & 13.43$\pm$0.03 & 12.40$\pm$0.03 & 11.65$\pm$0.03 \\
2000 May 13 & 275.367350 & +14.233563 & 13.50$\pm$0.05 & 12.39$\pm$0.04 & 11.64$\pm$0.04 \\
\cutinhead{2MASS J21481628+4003593,
$\mu$=1${\farcs}$33$\pm$0${\farcs}$24 yr$^{-1}$ at
$\theta$=75.6$^{\circ}\pm$0.3$^{\circ}$, $\Delta$t=0.46 yr}
1999 Dec 14 & 327.067843 & +40.066483 & 14.26$\pm$0.04 & 12.80$\pm$0.04 & 11.78$\pm$0.03 \\
2000 May 30\tablenotemark{a} & 327.068069 & +40.066525 & 14.15$\pm$0.04 & 12.78$\pm$0.04 & 11.77$\pm$0.04 \\
\enddata
\tablenotetext{a}{Entries in the 2MASS Point Source Catalog.}
\end{deluxetable} 

\begin{deluxetable}{llllllll}
\tabletypesize{\scriptsize}
\tablewidth{7.3in}
\tablecaption{Observations Log}
\tablehead{ 
\colhead{Object\tablenotemark{a}} & \colhead{Tel/Instmt} &
 \colhead{$\lambda$ ($\mu$m)} & \colhead{${\lambda}/{\Delta}{\lambda}$} & 
 \colhead{UT Date} & \colhead{z} & 
 \colhead{N~$\times$~t(s)\tablenotemark{b}} & \colhead{Calibrator}}
\startdata
2MASS J02052940$-$1159296 & Subaru/FOCAS & 0.585--1.025 & 1000 & 
  2007 Aug 20 (UT) & 1.2 & 2~$\times$~1200 & \nodata \\
\hline
2MASS J15074769$-$1627386 & Subaru/FOCAS & 0.585--1.025 & 1000 & 
  2007 Aug 21 (UT) & 1.5 & 2~$\times$~600 & \nodata \\
\hline
2MASS J18212815+1414010 & Subaru/FOCAS & 0.585--1.025 & 1000 & 
  2007 Aug 20 (UT) & 1.0 & 2~$\times$~1200 & SAO 10356 (G0 V) \\
2MASS J18212815+1414010 & Keck/NIRSPEC & 1.14--1.36 & 2500 & 
  2005 Jul 18 (UT) & 1.0 & 4~$\times$~300 & HD 165029 (A0 V) \\
2MASS J18212815+1414010 & IRTF/SpeX & 0.7--2.5 & 100 & 
  2005 Aug 10 (UT) & 1.0 & 8~$\times$~120 & HD 165029 (A0 V) \\
2MASS J18212815+1414010 & IRTF/SpeX & 0.9--2.4 & 1200 & 
  2005 Aug 10 (UT) & 1.1 & 12~$\times$~300 & HD 165029 (A0 V) \\
2MASS J18212815+1414010 & IRTF/SpeX & 1.9--4.2 & 1500 &
  2007 Jul 28 (UT) & 1.3 & 40~$\times$~30 & HD 165029 (A0 V) \\
2MASS J18212815+1414010 & IRTF/SpeX & 1.9--4.2 & 1500 &
  2007 Jul 28 (UT) & 1.5 & 40~$\times$~30 & HD 165029 (A0 V) \\
2MASS J18212815+1414010 & Spitzer/IRS & 5.2--15.3 & 90 &
  AOR 16201984 & \nodata & 3904 & \nodata \\
\hline
2MASS J21321145+1341584 & Subaru/FOCAS & 0.585--1.025 & 1000 & 
  2007 Aug 21 (UT) & 1.2 & 2~$\times$~1200 & \nodata \\
\hline
2MASS J21481628+4003593 & Subaru/FOCAS & 0.585--1.025 & 1000 & 
  2007 Aug 20 (UT) & 1.2 & 3~$\times$~1200 & SAO 207827 (G0 V) \\
2MASS J21481628+4003593 & Subaru/FOCAS & 0.585--1.025 & 1000 &
  2007 Aug 21 (UT) & 1.1 & 2~$\times$~1200 & SAO 207827 (G0 V) \\
2MASS J21481628+4003593 & Keck/NIRSPEC & 1.14--1.36 & 2500 & 
  2005 Dec 09 (UT) & 1.2 & 4~$\times$~300 & SAO 71693 (A0 V) \\
2MASS J21481628+4003593 & Keck/NIRSPEC & 0.94--1.15 & 2500 & 
  2006 Aug 04 (UT) & 1.2 & 4~$\times$~200 & HD 207220 (A0 V) \\
2MASS J21481628+4003593 & Keck/NIRSPEC & 1.10--1.31 & 2500 & 
  2006 Aug 04 (UT) & 1.2 & 4~$\times$~200 & HD 207220 (A0 V) \\
2MASS J21481628+4003593 & IRTF/SpeX & 0.7--2.5 & 100 &
  2005 Sep 08 (UT) & 1.1 & 8~$\times$~120 & HD 207220 (A0 V) \\
2MASS J21481628+4003593 & IRTF/SpeX & 0.7--2.5 & 100 &
  2005 Sep 09 (UT) & 1.1 & 8~$\times$~120 & HD 207220 (A0 V) \\
2MASS J21481628+4003593 & IRTF/SpeX & 0.9--2.4 & 1200 &
  2005 Sep 09 (UT) & 1.1 & 24~$\times$~120 & HD 207220 (A0 V) \\
2MASS J21481628+4003593 & IRTF/SpeX & 1.9--4.2 & 1500 &
  2006 Sep 01 (UT) & 1.1 & 40~$\times$~30 & HD 209932 (A0 V) \\
2MASS J21481628+4003593 & IRTF/SpeX & 1.9--4.2 & 1500 &
  2006 Sep 12 (UT) & 1.1 & 40~$\times$~30 & HD 209932 (A0 V) \\
2MASS J21481628+4003593 & IRTF/SpeX & 1.9--4.2 & 1500 &
  2006 Sep 12 (UT) & 1.1 & 40~$\times$~30 & HD 209932 (A0 V) \\
2MASS J21481628+4003593 & IRTF/SpeX & 1.9--4.2 & 1500 &
  2006 Sep 12 (UT) & 1.1 & 40~$\times$~30 & HD 209932 (A0 V) \\
2MASS J21481628+4003593 & IRTF/SpeX & 1.9--4.2 & 1500 &
  2007 Nov 11 (UT) & 1.1 & 40~$\times$~30 & HD 13936 (A0 V) \\
2MASS J21481628+4003593 & IRTF/SpeX & 1.9--4.2 & 1500 &
  2007 Nov 11 (UT) & 1.1 & 40~$\times$~30 & HD 13936 (A0 V) \\ 
2MASS J21481628+4003593 & IRTF/SpeX & 1.9--4.2 & 1500 &
  2007 Nov 11 (UT) & 1.2 & 40~$\times$~30 & HD 13936 (A0 V) \\ 
2MASS J21481628+4003593 & IRTF/SpeX & 1.9--4.2 & 1500 &
  2007 Nov 11 (UT) & 1.2 & 40~$\times$~30 & HD 13936 (A0 V) \\ 
2MASS J21481628+4003593 & IRTF/SpeX & 1.9--4.2 & 1500 &
  2007 Nov 11 (UT) & 1.3 & 40~$\times$~30 & HD 209932 (A0 V) \\
2MASS J21481628+4003593 & Spitzer/IRS & 5.2--15.3 & 90 &
  AOR 16176128 & \nodata & 3904 & \nodata \\
\hline
2MASS J22443167+2043433 & IRTF/SpeX & 0.7--2.5 & 100 & 
  2005 Sep 09 (UT) & 1.1 & 12~$\times$~120 & HD 207720 (A0 V) \\
\enddata 
\tablenotetext{a}{J2000 coordinates from the 2MASS All-Sky Point Source Catalog.}
\tablenotetext{b}{Number of integrations times integration time.}
\end{deluxetable}

\begin{deluxetable}{lccccccc}
\tabletypesize{\scriptsize}
\tablewidth{6.9in}
\tablecaption{L Dwarfs within 10 Parsecs}
\tablehead{ 
\colhead{} & \colhead{} & \colhead{} & \colhead{} &
\multicolumn{3}{c}{Distance (pc)} & \colhead{} \\
\cline{5-7} \\
\colhead{Name of} & \colhead{Discovery} & \colhead{Optical Sp.} & 
\colhead{$J$\tablenotemark{b}} &  
\colhead{from} & \colhead{from} & \colhead{Adopted} & \colhead{Distance}\\
\colhead{L Dwarf} & \colhead{Reference} & \colhead{Type\tablenotemark{a}} & 
\colhead{(mag)} & \colhead{D$_\mathrm{est}$\tablenotemark{c,d}} & 
\colhead{$\pi_{trig}$} & \colhead{Distance\tablenotemark{e}} &
\colhead{Reference}} 
\startdata
DENIS-P J0255$-$4700 & 1 & L8 & 13.25~$\pm$~0.03 &
 5.1~$\pm$~0.7 & 5.0~$\pm$~0.1 & 5.0~$\pm$~0.1 & 9 \\
2MASSW J1507476$-$162738 & 2 & L5 & 12.83~$\pm$~0.03 &
 6.6~$\pm$~0.9 & 7.33~$\pm$~0.03 & 7.33~$\pm$~0.03 & 10 \\
2MASS J17502484$-$0016151 & 3 & L5.5\tablenotemark{f} & 
 13.29~$\pm$~0.02 & 7.3~$\pm$~1.0 & \nodata & 7.3~$\pm$~1.0 & 8 \\
2MASSI J0835425$-$081923 & 4 & L5 & 13.17~$\pm$~0.02 &
 7.7~$\pm$~1.0 & \nodata & 7.7~$\pm$~1.0 & 8 \\
2MASSW J0036159+182110 & 2 & L3.5 & 12.47~$\pm$~0.03 &
 8.0~$\pm$~1.1 & 8.76~$\pm$~0.06 & 8.76~$\pm$~0.06 & 10 \\
2MASS J03552337+1133437 & 5 & L6: & 14.05~$\pm$~0.02 & 
 9.4~$\pm$~1.3 & \nodata & 9.4~$\pm$~1.3 & 8 \\
2MASSW J1515008+484742 & 6 & L6 & 14.11~$\pm$~0.03 & 
 9.7~$\pm$~1.3 & \nodata & 9.7~$\pm$~1.3 & 8 \\
2MASS J02572581$-$3105523 & 7 & L8 & 14.67~$\pm$~0.04 &
 9.7~$\pm$~1.3 & \nodata & 9.7~$\pm$~1.3 & 8 \\
2MASS J18212815+1414010 & 8 & L4.5 & 13.43~$\pm$~0.03 & 
 9.8~$\pm$~1.3 & \nodata & 9.8~$\pm$~1.3 & 8 \\
2MASS J21481628+4003593 & 8 & L6 & 14.15~$\pm$~0.04 & 
 9.9~$\pm$~1.3 & \nodata & 9.9~$\pm$~1.3 & 8 \\
\enddata 
\tablenotetext{a}{Optical spectral types, unless otherwise noted.}
\tablenotetext{b}{Photometry from 2MASS All-Sky Point Source Catalog.}
\tablenotetext{c}{Spectrophotometric distance estimate -- see $\S$3.4.}
\tablenotetext{d}{D$_\mathrm{est}~=~10^{(m_J-M_J)/5}$,
$\sigma_D$~=~ln(10)$/$5~$\times$~D$_{est}~\times~\sqrt{\sigma_{m_J}^2
+ \sigma_{M_J}^2}$.}
\tablenotetext{e}{Distance adopted from trigonometric parallax if measured.}
\tablenotetext{f}{Near-infrared spectral type.}
\tablecomments{Discovery and Distance References: (1) \cite{martin1999},
 (2) \cite{reid2000}, (3) \cite{kendall2007}, (4) \cite{cruz2003}, 
(5) Reid et al$.$, in prep, (6) \cite{wilson2003}, 
(7) Kirkpatrick et al$.$, in prep, (8) this paper, 
(9) \cite{costa2006}, (10) \cite{dahn2002}.}
\end{deluxetable}

\begin{deluxetable}{lll} 
\tablewidth{4.5in}
\tablecaption{Properties of the Two New L Dwarfs}
\tablehead{ 
\colhead{Parameter} & \colhead{2MASS J1821+1414} & 
\colhead{2MASS J2148+4003}} 
\startdata
J$-K_s$\tablenotemark{a} & 1.78~$\pm$~0.04 & 2.38~$\pm$~0.06 \\
J$-$H\tablenotemark{a} & 1.03~$\pm$~0.04 & 1.37~$\pm$~0.06 \\
H$-K_s$\tablenotemark{a} & 0.75~$\pm$~0.04 & 1.01~$\pm$~0.06 \\
Opt SpT & L4.5 & L6 \\
NIR SpT & L5 pec & L6.5 pec \\
d$_\mathrm{est}$ (pc) & 9.8~$\pm$~1.3\tablenotemark{b}~~(9.8~$\pm$~1.5)\tablenotemark{c} & 
  9.9~$\pm$~1.3\tablenotemark{b}~~(7.9~$\pm$~1.2)\tablenotemark{c}\\
V$_\mathrm{tan}$ (km s$^{-1}$) & 9.8~$\pm$~1.4 & 62.4~$\pm$~3.3 (49.8~$\pm$~3.1)\tablenotemark{d}\\
\cutinhead{Bolometric Measurements/Estimates}
Log$_{10}$(L$_{bol}$/L$_\odot$)\tablenotemark{e} & $-$4.12~$\pm$~0.12 &
  $-$4.07~$\pm$~0.12 \\
m$_{bol}$\tablenotemark{f} (mag) & 15.01~$\pm$~0.01 & 14.91~$\pm$~0.01 \\
M$_{bol}$ (mag) & 15.05~$\pm$~0.29 & 14.93~$\pm$~0.29 \\
f$_{bol}$\tablenotemark{f}~~(W m$^{-2}$) & (2.51~$\pm$~0.03)~e$-$14 & (2.77~$\pm$~0.04)~e$-$14 \\
\enddata 
\tablenotetext{a}{Colors are computed from the values listed in the
2MASS Point Source Catalog, see Table 1.}
\tablenotetext{b}{Distance estimate is from the optical spectral type versus 
M$_J$ relation derived by \cite{looper2008}.}
\tablenotetext{c}{Distance estimate is from the optical spectral type versus 
M$_{K_s}$ relation derived by \cite{looper2008}.}
\tablenotetext{d}{Based on a distance estimate of 7.9 pc.}
\tablenotetext{e}{Log$_{10}$(L$_{bol}$/L$_\odot$)~=~(M$_{bol~\odot} -$
M$_{bol}$)~$\times$~2/5, where M$_{bol~\odot}$~=~+4.76.}
\tablenotetext{f}{Values for m$_{bol}$ and f$_{bol}$ are independent of distance.}
\end{deluxetable} 

\clearpage

\begin{thebibliography}{}
\bibitem[Allers et al.(2007)]{allers2007} Allers, K.~N., et al.\ 
  2007, \apj, 657, 511
\bibitem[Barnard(1916)]{barnard1916} Barnard, E.~E.\ 1916, \aj, 29, 181.
\bibitem[Borysow et al.(1997)]{borysow1997} Borysow, ~A., Jorgensen,
  U.~G., Zheng, ~C.\ 1997, A\&A, 324, 185.
\bibitem[Burgasser et al.(2008)]{burgasser2008} Burgasser, A.~J., 
  Looper, D.~L., Kirkpatrick, J.~D., Cruz, K.~L., 
  \& Swift, B.~J.\ 2008, \apj, 674, 451 
\bibitem[Burgasser(2007)]{burgasser2007b} Burgasser, A.~J.\ 2007, 
  \apj, 659, 655 
\bibitem[Burgasser et al.(2007)]{burgasser2007a} Burgasser, A.~J., 
  Looper, D.~L., Kirkpatrick, J.~D., \& Liu, M.~C.\ 2007, \apj, 658, 557 
\bibitem[Burgasser(2004)]{burgasser2004} Burgasser, A.~J.\ 2004, 
  \apjl, 614, L73 
\bibitem[Burgasser et al.(2003)]{burgasser2003} Burgasser, A.~J., 
  et al.\ 2003, \apj, 592, 1186.
\bibitem[Costa et al.(2006)]{costa2006} Costa, Edgardo, M\'{e}ndez,
  R.~A., et al.\ 2006, AJ, 132, 1234.
\bibitem[Cruz et al.(2007)]{cruz2007} Cruz, K.~L., et al.\ 2007, 
  \aj, 133, 439
\bibitem[Cruz et al.(2003)]{cruz2003} Cruz, K.~L., Reid, I.~N., Liebert,
  J., Kirkpatrick, J.~D., \& Lowrance, P.~J.\ 2003, \aj, 126, 2421.
\bibitem[Cushing et al.(2006)]{cushing2006} Cushing, M.~C., et al.\
  2006, \apj, 648, 614. 
\bibitem[Cushing et al.(2005)]{cushing2005} Cushing, M.~C., Rayner,
  J.~T., \& Vacca, W.~D.\ 2005, \apj, 623, 1115.
\bibitem[Cushing et al.(2004)]{cushing2004} Cushing, M.~C., Vacca, 
  W.~D., \& Rayner, J.~T.\ 2004, \pasp, 116, 362. 
\bibitem[Dahn et al.(2002)]{dahn2002} Dahn, C.~C., et al.\ 2002, \aj,
  124, 1170.
\bibitem[Deacon \& Hambly(2007)]{deacon2007} Deacon, N.~R., \& 
  Hambly, N.~C.\ 2007, \aap, 468, 163 
\bibitem[Deacon et al.(2005)]{deacon2005} Deacon, N.~R., 
  Hambly, N.~C., \& Cooke, J.~A.\ 2005, \aap, 435, 363 
\bibitem[Epchtein et al.(1997)]{epchtein1997} Epchtein, N., et al.\ 
  1997, The Messenger, 87, 27 
\bibitem[Finch et al.(2007)]{finch2007} Finch, C.~T., Henry, T.~J., 
  Subasavage, J.~P., Jao, W.-C., \& Hambly, N.~C.\ 2007, \aj, 133, 2898 
\bibitem[Giclas et al.(1978)]{giclas1978} Giclas, H.~L., Burnham, 
  R., \& Thomas, N.~G.\ 1978, Lowell Observatory Bulletin, 8, 89.
\bibitem[Hamuy et al.(1994)]{1994PASP..106..566H} Hamuy, M., Suntzeff, 
  N.~B., Heathcote, S.~R., Walker, A.~R., Gigoux, P., 
  \& Phillips, M.~M.\ 1994, \pasp, 106, 566 
\bibitem[Houck et al.(2004)]{houck2004} Houck, J.~R., et al.\ 
  2004, \apjs, 154, 18 
\bibitem[Innes(1915)]{innes1915} Innes, R.~T.~.A.\ 1915, Union Observatory 
  Circular, No.\ 30.
\bibitem[Kashikawa et al.(2002)]{2002PASJ...54..819K} Kashikawa, N., et 
  al.\ 2002, \pasj, 54, 819 
\bibitem[Kendall et al.(2007)]{kendall2007} Kendall, T.~R., Jones, 
  H.~R.~A., Pinfield, D.~J., Pokorny, R.~S., Folkes, S., Weights, D., 
  Jenkins, J.~S., \& Mauron, N.\ 2007, \mnras, 374, 445 
\bibitem[Kirkpatrick et al.(2006)]{kirkpatrick2006} Kirkpatrick,
  J.~D., et al.\ 2006, \apj, 639, 1120.
\bibitem[Kirkpatrick(2005)]{kirkpatrick2005} Kirkpatrick, J.~D.\
  2005, \araa, 43, 195. 
\bibitem[Kirkpatrick et al.(2000)]{kirkpatrick2000} Kirkpatrick, J.~D., 
  et al.\ 2000, \aj, 120, 447 
\bibitem[Kirkpatrick et al.(1999)]{kirkpatrick1999} Kirkpatrick, J.~D.,
  Reid, I.~N., Liebert, ~J., et al.\ 1999, \apj, 519, 802.
\bibitem[Lepine et al.(2008)]{lepine2008b} Lepine, S., Shara, 
  M.~M., Rich, R.~M., Wittenberg, A., Halmo, M., 
  \& Bongiorno, B.\ 2008, ArXiv e-prints, 805, arXiv:0805.4736 
\bibitem[L{\'e}pine(2008)]{lepine2008a} L{\'e}pine, S.\ 2008, \aj, 
  135, 2177 
\bibitem[L{\'e}pine et al.(2002)]{lepine2002} L{\'e}pine, S., 
  Shara, M.~M., \& Rich, R.~M.\ 2002, \aj, 124, 1190 
\bibitem[Looper et al.(2008)]{looper2008} Looper, D.~L., Gelino, 
  C.~R., Burgasser, A.~J., 
  \& Kirkpatrick, J.~D.\ 2008, ArXiv e-prints, 803, arXiv:0803.0544 
\bibitem[Looper et al.(2007b)]{looper2007b} Looper, D.~L., Burgasser, 
  A.~J., Kirkpatrick, J.~D., \& Swift, B.~J.\ 2007b, \apjl, 669, L97 
\bibitem[Looper et al.(2007a)]{looper2007a} Looper, D.~L., 
  Kirkpatrick, J.~D., \& Burgasser, A.~J.\ 2007a, \aj, 134, 1162 
\bibitem[Lucas et al.(2001)]{lucas2001} Lucas, P.~W., Roche, 
  P.~F., Allard, F., \& Hauschildt, P.~H.\ 2001, \mnras, 326, 695 
\bibitem[Luyten(1979)]{luyten1979} Luyten, W.\ 1979, Minneapolis: 
  University of Minnesota, 1979, NLTT Catalog, Vol.\ I-II.
\bibitem[Luyten(1980)]{luyten1980} Luyten, W.\ 1980, Minneapolis: 
  University of Minnesota, 1980, NLTT Catalog, Vol.\ III-IV.
\bibitem[Mart\'{i}n et al.(1999)]{martin1999} Mart\'{i}n, E.~L., 
  Delfosse, X., et al.\ 1999, \aj, 118, 2466.
\bibitem[Massey \& Gronwall(1990)]{1990ApJ...358..344M} Massey, P., 
  \& Gronwall, C.\ 1990, \apj, 358, 344
\bibitem[McGovern et al.(2004)]{mcgovern2004} McGovern, M.~R., 
  Kirkpatrick, J.~D., McLean, I.~S., Burgasser, A.~J., Prato, L., 
  \& Lowrance, P.~J.\ 2004, \apj, 600, 1020 
\bibitem[McLean et al.(2003)]{mclean2003} McLean, I.~S., McGovern, 
  M.~R., Burgasser, A.~J., Kirkpatrick, J.~D., Prato, L., \& Kim, S.~S.\ 
  2003, \apj, 596, 561.
\bibitem[McLean et al.(2000)]{mclean2000} McLean, I.~S., Graham, 
  J.~R., Becklin, E.~E., Figer, D.~F., Larkin, J.~E., Levenson, N.~A., \& 
  Teplitz, H.~I.\ 2000, \procspie, 4008, 1048.
\bibitem[McLean et al.(1998)]{mclean1998} McLean, I.~S., et al.\ 
  1998, \procspie, 3354, 566.
\bibitem[Monet et al.(2003)]{monet2003} Monet, D.~G., et al.\ 2003, \aj,
  125, 984.
\bibitem[Pokorny et al.(2004)]{pokorny2004} Pokorny, R.~S., 
  Jones, H.~R.~A., Hambly, N.~C., \& Pinfield, D.~J.\ 2004, \aap, 421, 763 
\bibitem[Rayner et al.(2003)]{rayner2003} Rayner, J.~T., Toomey, 
  D.~W., Onaka, P.~M., Denault, A.~J., Stahlberger, W.~E., Vacca, W.~D., 
  Cushing, M.~C., \& Wang, S.\ 2003, \pasp, 115, 362.
\bibitem[Reid et al.(2000)]{reid2000} Reid, I.~N., Kirkpatrick, J.~D.,
  Gizis, J.~E., Dahn, C.~C., Monet, D.~G., Williams, R.~J., Liebert, J.,
  \& Burgasser, A.~J.\ 2000, \aj, 119, 369.
\bibitem[Ross(1939)]{ross1939} Ross, F.~E.\ 1939, \aj, 48, 163.
\bibitem[Skrutskie et al.(2006)]{skrutskie2006} Skrutskie, M.~F., et 
  al.\ 2006, \aj, 131, 1163 
\bibitem[Snowden et al.(1998)]{snowden1998} Snowden, S.~L., Egger, 
  R., Finkbeiner, D.~P., Freyberg, M.~J., \& Plucinsky, P.~P.\ 1998, \apj, 493, 715 
\bibitem[Vacca et al.(2003)]{vacca2003} Vacca, W.~D., Cushing, 
  M.~C., \& Rayner, J.~T.\ 2003, \pasp, 115, 389. 
\bibitem[Vrba et al.(2004)]{vrba2004} Vrba, F.~J., Henden, A.~A.,
  Luginbuhl, C.~B., Guetter, H.~H., Munn, J.~A.\ 2004, \aj, 127, 2948.
\bibitem[Werner et al.(2004)]{werner2004} Werner, M.~W., et al.\ 
  2004, \apjs, 154, 1 
\bibitem[Wilson et al.(2003)]{wilson2003} Wilson, J.~C., Miller,
  N.~A., Gizis, J.~E., Skrutskie, M.~F., Houck, J.~R., Kirkpatrick, J.~D.,
  Burgasser, A.~J., \& Monet, D.~G.\ 2003, Proceedings of IAU Symposium
  \#211, 211, 197
\bibitem[Wolf(1918)]{wolf1918} Wolf, M.\ 1918, Astronomische 
  Nachrichten, 206, 237 
\bibitem[York et al.(2000)]{york2000} York, D.~G., et al.\ 2000, 
  \aj, 120, 1579 
\end{thebibliography}
 \end{document}